\PassOptionsToPackage{hypertexnames=false}{hyperref}
\documentclass[pdflatex,iicol,sn-mathphys-num]{sn-jnl}

\usepackage{amsmath,amssymb,bm}
\usepackage{graphicx}
\usepackage{booktabs,tabularx,array}
\usepackage{float}
\usepackage{placeins}
\usepackage{xcolor}
\usepackage{xurl}
\providecommand{\burl}[1]{\url{#1}}
\graphicspath{{figures/}}
\hypersetup{pdfauthor={Ruixin Zhou and Boliang Yu}}

\begin{document}

\title[Microscopic decorrelation and collective ordering]{Rapid microscopic decorrelation precedes sustained collective ordering in quenched two-dimensional colloids}
\author[1]{\fnm{Ruixin} \sur{Zhou}}
\author*[1]{\fnm{Boliang} \sur{Yu}}
\email{yuboliang@sjtu.edu.cn}
\affil[1]{\orgdiv{School of Physics and Astronomy},
\orgname{Shanghai Jiao Tong University},
\orgaddress{\city{Shanghai}, \postcode{200240}, \country{China}}}

\abstract{Independent Brownian noise can rapidly separate trajectories launched from the same microscopic configuration, but configuration-space distance need not reveal their later assembly outcomes. We studied this distinction in 480 two-dimensional Brownian-dynamics trajectories generated as three independent-noise replicas of 40 parent configurations under four finite-duration quenches. Ninety-two of 160 replica sets sharing one parent and one protocol reached both ordered and non-ordered terminal outcomes. A relative-diffusion calculation relates the free-particle mean-squared replica separation to the time-integrated diffusivity. The measured mean-squared separation remained about one-sixth of the free-particle reference, while its scaling was indistinguishable for replica pairs with the same and different outcomes. In contrast, outcome-resolved global sixfold order, ordered-particle population and coherence, and reciprocal-space translational order separated progressively after quench completion. With quench durations weighted equally, global order, the state score combining global and population order, coherence, and translational order remained informative 0.50 reduced time units before terminal assessment. This result persisted after either changing the terminal window or reducing the integration timestep. Transient threshold crossings were common, and 94.5\% of non-ordered crossings subsequently lost the required ordered-particle population. An independent $N=128$ ensemble showed state-score discrimination at temperatures for which both terminal classes occurred. The assembly dynamics therefore separate generic microscopic decorrelation from the later emergence and finite-window persistence of collective order associated with terminal structure.}


\maketitle

\section{Introduction}

Nonequilibrium colloidal assembly is shaped jointly by particle interactions, thermal fluctuations, and the history of the driving protocol \citep{Whitesides2002SelfAssembly,Glotzer2007Anisotropy,Whitelam2015DynamicPathways,Grant2011AssemblyMechanisms}. Even replicas prepared in the same microscopic configuration can reach different structures because their thermal-noise histories branch. This stochasticity motivates a dynamical question: as independent noise removes microscopic similarity, which components of the resulting motion are relevant to the later assembly outcome \citep{Auer2001NucleationRate,Gasser2001RealSpaceNucleation,Royall2024HardSpheres}?

An isoconfigurational ensemble, formed by launching independent-noise replicas from a shared parent, separates the common initial structure from subsequent stochastic forcing \citep{WidmerCooper2004Reproducibility}. It also makes two complementary comparisons possible. At the pair level, one can ask whether the magnitude of coordinate or neighbor-network separation distinguishes replicas that later share an outcome from replicas that reach different outcomes. At the trajectory level, one can instead ask when interpretable collective coordinates become associated with the later outcomes. The first comparison probes generic decorrelation; the second probes motion along candidate reaction coordinates \citep{Bolhuis2002TPS,E2006TransitionPaths}. Their relation is nontrivial because a high-dimensional distance accumulates displacements in many directions, whereas assembly depends on a restricted set of collective modes.

Two-dimensional colloids offer experimentally accessible collective coordinates for this purpose. Sixfold order can be separated into global alignment, the population of locally ordered particles, and coherence or connectivity within ordered regions \citep{Halperin1978Melting,Zahn1999TwoStage,Bernard2011TwoStep,Kapfer2015Melting}. Bond-orientational order is a well-established crystallization coordinate \citep{Kawasaki2010CrystalNucleus,Russo2012MicroscopicPathway}, although local magnitude alone does not capture the full structure \citep{Mickel2013BondOrderShortcomings}. We therefore compare orientational measures with a reciprocal-space translational-order score derived from the first reciprocal shell. Their joint evolution tests whether discrimination between terminal outcomes reflects collective structural ordering beyond the thresholds used to define the terminal state.

Here we analyze 480 Brownian-dynamics trajectories at \ensuremath{N=64}: three independent-noise replicas of 40 parents under each of four quench durations. A relative-diffusion reference derived from the Langevin equation sets the diffusive scale for replica separation, while a finite-time outcome-probability decomposition separates future-noise branching from parent-to-parent heterogeneity. Shared parents frequently reach both terminal outcomes, and coordinate and neighbor-graph distances grow without distinguishing the two pair types. Outcome-resolved orientational and translational coordinates instead separate progressively after quench completion. Transient crossings further distinguish short-lived alignment from sustained ordered-particle growth. The resulting picture is a separation between generic noise-driven decorrelation and the slower development and persistence of outcome-associated collective order.

\section{Model and methods}

\subsection{Dynamics, parent ensemble, and quench protocol}

We simulated particles in a two-dimensional square periodic box at number density \ensuremath{\rho=0.82}. The primary ensemble used \ensuremath{N=64} and box length \ensuremath{L=(N/\rho)^{1/2}=8.8345}. Particles interacted through a truncated-and-shifted Lennard–Jones potential,

\begin{equation}
  U(r)=
  \begin{cases}
    U_{\mathrm{LJ}}(r)-U_{\mathrm{LJ}}(r_c), & r<r_c,\\
    0, & r\ge r_c,
  \end{cases}
\end{equation}
with
\begin{equation}
  U_{\mathrm{LJ}}(r)=4\varepsilon
  \left[(\sigma/r)^{12}-(\sigma/r)^6\right].
\end{equation}

The cutoff was \ensuremath{r_c=2.5\sigma}. Reduced units were used, with \ensuremath{\sigma=\varepsilon=\mu=k_{\mathrm B}=1}, where \ensuremath{\mu} is the mobility. The reduced time unit is \ensuremath{t_0=\sigma^2/(\mu\varepsilon)}. At \ensuremath{T_{\mathrm{low}}=0.50}, the free-particle diffusion coefficient is \ensuremath{D=\mu k_{\mathrm B}T=0.5} and the Brownian diffusion time \ensuremath{\tau_B=\sigma^2/D=2t_0}. Periodic forces used minimum-image separations. For the primary box, \ensuremath{L/2=4.417>r_c}, and no long-range tail correction was applied.

Large-system equilibrium simulations of the same range-limited \ensuremath{r_c=2.5\sigma} interaction place \ensuremath{T=0.50}, \ensuremath{\rho=0.82} close to the reported low-temperature solid--fluid coexistence region and show that the phase diagram depends quantitatively on the cutoff \citep{Li2020PhaseBehavior2DLJ}. This phase diagram provides equilibrium context for the present finite quenches. Because the \ensuremath{N=64} square periodic cell is incommensurate with a defect-free triangular tiling, ``ordered'' and ``non-ordered'' below denote finite-window structural outcomes rather than equilibrium phases.

Following standard Brownian-dynamics constructions \citep{Ermak1978Brownian,Branka1998BrownianAlgorithms}, we integrated the overdamped Langevin equation by Euler–Maruyama,

\begin{equation}
  \mathbf r_i(t+\Delta t)=\mathbf r_i(t)+\mu \mathbf F_i(t)\Delta t+[2\mu T(t)\Delta t]^{1/2} \boldsymbol{\xi}_i(t),
\end{equation}

where \ensuremath{\mathbf r_i} and \ensuremath{\mathbf F_i} are two-dimensional vectors and the Cartesian components of \ensuremath{\xi_i} are independent standard normal variates. The scalar mobility corresponds to free-draining Brownian dynamics; hydrodynamic interactions were omitted. The primary ensemble used \ensuremath{\Delta t=5\times10^{-5}}. Numerical implementation and trajectory-storage details are given in Supplementary Information.

The Langevin equation also supplies a reference scale for the separation of two independent-noise replicas. For replicas $a$ and $b$, let $\delta\mathbf r_i=\mathbf r_i^{(a)}-\mathbf r_i^{(b)}$ and $D(t)=\mu k_{\rm B}T(t)$. Their relative motion obeys
\begin{equation}
 d\delta\mathbf r_i
 =\mu\left(\mathbf F_i^{(a)}-\mathbf F_i^{(b)}\right)dt
 +\sqrt{4D(t)}\,d\mathbf W_i.
 \label{eq:relative-motion}
\end{equation}
At the branching time $t_b$, the coordinates and forces are identical. Defining the translation-corrected mean-square separation by $D_{\rm RMS}^2=N^{-1}\sum_i|\delta\mathbf r_i-\overline{\delta\mathbf r}|^2$, the free-particle reference in two dimensions is
\begin{equation}
 \left\langle D_{\rm RMS}^2(t)\right\rangle_{\rm free}
 =8\left(1-\frac{1}{N}\right)
 \int_{t_b}^{t}D(s)\,ds .
 \label{eq:free-separation}
\end{equation}
Interactions subsequently generate force differences that modify the free relative diffusion. We use Eq.~\eqref{eq:free-separation} as a reference rather than as an interacting-particle fit.

Each primary parent was initialized from a weakly perturbed square arrangement and evolved independently for 2.0 reduced time units at \ensuremath{T_{\mathrm{high}}}=1.8. The 40 parents used distinct initialization and Brownian seeds. We therefore refer to them as independently generated high-temperature parent states, with the finite burn-in defining the sampling scope.

After parent generation, temperature was held at \ensuremath{T_{\mathrm{high}}} for \ensuremath{t_{\mathrm{hold}}}=0.2 and then decreased linearly to \ensuremath{T_{\mathrm{low}}}=0.50 over a duration \ensuremath{\tau_q}. The protocol was

\begin{equation}
  \begin{aligned}
  T(t)&=T_{\mathrm{high}},
  &&t<t_{\mathrm{hold}},\\
  T(t)&=T_{\mathrm{high}}
  -\frac{T_{\mathrm{high}}-T_{\mathrm{low}}}{\tau_q}\\[-1mm]
  &\quad\times(t-t_{\mathrm{hold}}),\\[-1mm]
  &\quad t_{\mathrm{hold}}\le t\le t_{\mathrm{hold}}+\tau_q.
  \end{aligned}
\end{equation}

and \ensuremath{T(t)=T_{\mathrm{low}}} thereafter. We used \ensuremath{\tau_q}=0.15, 0.60, 1.50, and 3.00. Every trajectory continued for 2.0 reduced time units after the quench ended. The quantity \ensuremath{\tau_q} is the quench duration, equivalently the protocol timescale.

For every parent and duration, three replicas reused the same stored pre-quench trajectory. At \ensuremath{t=t_{\mathrm{hold}}}, each replica switched to a distinct Brownian-noise seed. The primary data set therefore contained

\begin{equation}
  \begin{aligned}
  N_{\mathrm{traj}}
  &=40\ \mathrm{parents}\times4\ \mathrm{durations}\\
  &\quad\times3\ \mathrm{replicas}
  =480.
  \end{aligned}
\end{equation}

We refer to the three replicas sharing one parent and one quench duration as a replica set. If $q_{p,c}$ is the finite-time ordered-outcome probability for parent $p$ and protocol $c$, $p_c=E_p q_{p,c}$, and $d_c$ is the probability that two independent replicas have different outcomes, then
\begin{equation}
 p_c(1-p_c)
 =\frac{d_c}{2}+\operatorname{Var}_p(q_{p,c}).
 \label{eq:branch-decomposition}
\end{equation}
This identity follows from $d_c=2E_p[q_{p,c}(1-q_{p,c})]$. It separates within-parent future-noise branching from variation in finite-time outcome propensity among parents. Estimation with three replicas and complete-parent resampling is described in Supplementary Information.

The primary design is summarized in Table~\ref{tab:design}; all primary and comparison ensembles are listed in Supplementary Information, Table S1.

\subsection{Structural coordinates and finite-window outcomes}

Neighbors of particle i were defined by \ensuremath{r_{ij}}\ensuremath{\le}\ensuremath{r_{\mathrm{nb}}} with \ensuremath{r_{\mathrm{nb}}}=1.5. This fixed first-shell graph cutoff was chosen independently of terminal outcomes. The complex local sixfold bond-orientational order was defined in the standard form \citep{Steinhardt1983BondOrder} as

\begin{equation}
  \psi _{6,i}=n_i^{-1} \sum_{j\in \mathcal{N}_i} \exp(6 i \theta _{ij}),
\end{equation}

where \ensuremath{n_i} is the coordination number and \ensuremath{\theta_{ij}} is the bond angle. The global magnitude was

\begin{equation}
  \Psi _6=|N^{-1} \sum_i \psi _{6,i}|.
\end{equation}

A particle was assigned to the locally ordered mask O when \ensuremath{5\le n_i\le 7} and \ensuremath{|\psi_{6,i}|\ge 0.6}; its population fraction was \ensuremath{f_{\rm ord}=|O|/N}. This mask supplies the population definition used below and the connectivity measures in Supplementary Information \citep{Mickel2013BondOrderShortcomings}.

The ordered-particle orientational coherence was

\begin{equation}
  C_{\mathrm{ord}}=|\sum_{i\in O} \psi _{6,i}| / \sum_{i\in O}|\psi _{6,i}|,
\end{equation}

with \ensuremath{C_{\mathrm{ord}}}=0 for an empty ordered set. Additional neighbor-wide coherence and ordered-mask connectivity measures, together with normalization sensitivities, are defined in Supplementary Information.

To test whether the outcome association extended beyond bond-orientational coordinates, we evaluated reciprocal-space translational order through the static structure factor at reciprocal vectors allowed by the square periodic box,

\begin{align}
  S(\mathbf{k})
  &=N^{-1}\left|\sum_{j=1}^{N}
  \exp(i\mathbf{k}\cdot\mathbf{r}_j)\right|^2,\\
  \mathbf{k}&=\frac{2\pi}{L}(n_x,n_y).
\end{align}

For a triangular reference, \ensuremath{a_\triangle=[2/(\sqrt{3}\rho)]^{1/2}} and \ensuremath{k_0=4\pi/(\sqrt{3}a_\triangle)}. The translational-order score \ensuremath{s_T} was the mean of the three largest \ensuremath{S(\mathbf{k})} values in the first-shell annulus around \ensuremath{k_0}. The number three was chosen to match the number of inversion-unique directions in an ideal triangular first shell. Its definition uses neither bond angles, the ordered mask, terminal labels, nor fitted models. The reciprocal-grid construction and an annular-mean sensitivity are described in Supplementary Information.

Terminal statistics were calculated over the last 15\% of stored frames, with a minimum of five frames. A trajectory received the finite-window ordered label when all three conditions held within that window:

\begin{equation}
  \begin{aligned}
  \operatorname{mean}(\Psi _6)&\ge 0.55,\\
  \operatorname{mean}(f_{\mathrm{ord}})&\ge 0.75,\\
  \operatorname{sd}(\Psi _6)&\le 0.08.
  \end{aligned}
\end{equation}

Every other trajectory received the finite-window non-ordered label. These labels classify the structure reached by the end of the simulated two-unit hold and are used as terminal outcomes throughout the analysis.

These cutoffs define the primary terminal outcome and were held fixed during model training and the primary analyses; only the explicitly identified terminal-window and threshold-sensitivity analyses recomputed terminal outcomes. We also recorded the first time \ensuremath{t^*} at which a trailing window spanning 7.5\% of that trajectory, with a minimum of two stored frames, simultaneously crossed the \ensuremath{\Psi_6} and \ensuremath{f_{\rm ord}} thresholds. A finite crossing followed by a terminal non-ordered label was classified as transient. Alternative terminal windows, threshold sensitivities, and the complete crossing-time procedure are provided in Supplementary Information.

\subsection{Observation times, state score, and replica divergence}

We sampled structure at eight predefined observation phases: the shared pre-quench state; 10\%, 25\%, 50\%, and 100\% of the trajectory-specific quench; and absolute times 0.25, 0.50, and 1.00 after quench completion. Figures abbreviate these phases as pre, q10, q25, q50, q100, p025, p050, and p100. All observation histories ended before, and did not overlap, the terminal window.

To follow structural development across protocols, quench-relative coordinates aligned equal fractions of each quench, whereas absolute post-quench coordinates aligned equal elapsed times after quench completion. We write the latter as \ensuremath{t_{\rm post}=t-t_{\mathrm{hold}}-\tau_q}. The dense structural profiles used 21 quench-relative and 24 absolute post-quench coordinates, giving 45 sampled times in total. Frame selection, history construction, and window-separation checks are given in Supplementary Information.

The threshold-normalized structural score used the latest eligible frame:

\begin{equation}
  s_{\mathrm{state}}=\min(\Psi _6/0.55, f_{\mathrm{ord}}/0.75).
\end{equation}

Its denominators are the instantaneous structural thresholds in the terminal ordered-label rule. We retain the short name ``state score'' below; it is a summary relative to this operational outcome definition, not a new thermodynamic order parameter.

For two replicas from the same parent and duration, we measured particle-identity root-mean-square (RMS) separation using periodic minimum-image displacements after subtracting their mean translational displacement. We also formed radius graphs at \ensuremath{r_{\mathrm{nb}}}=1.5 and measured the Jaccard distance between undirected edge sets. The pair-level target was whether the two replicas had different terminal labels.

\subsection{Evaluation and uncertainty}

As a flexible representation control, a GraphSAGE encoder and long short-term memory network were also trained on local particle environments and their recent evolution \citep{Hamilton2017GraphSAGE,Hochreiter1997LSTM}. These graph models probe whether the tested local histories contain outcome information beyond the interpretable collective coordinates; their architectures, inputs, and optimization settings are provided in Supplementary Information.

The graph models were evaluated by parent-grouped cross-validation, with all 12 descendants of a parent kept in the same fold and model selection confined to training and validation data \citep{Roberts2017CrossValidation,Varma2006CVBias}. This procedure yielded held-out scores for every trajectory. Outcome discrimination for both structural coordinates and learned scores was quantified by the area under the receiver operating characteristic curve (AUROC), with equal weight assigned to the four quench durations.

Uncertainty was estimated by resampling complete parent clusters \citep{Efron1979Bootstrap,Field2007ClusterBootstrap}. For learned models, a hierarchical interval additionally propagated variation across repeated fits. Pointwise curves describe structural development, and simultaneous bands account for comparison over sampled times or observables. Fixed-lead measurements at 0.25, 0.50, 0.75, and 1.00 before the terminal-window start hold the distance to terminal assessment constant across protocols. For duration-adjusted correlations, ranks were centered within each quench duration before pooling. Between-parent outcome variation was corrected for the finite-replica Bernoulli contribution. Full resampling procedures and association statistics are given in Supplementary Information.

\subsection{Timestep and larger-system comparisons}

To test sensitivity to time discretization, we repeated the primary design with the same 40 parents, four durations, and three replica identifiers at \ensuremath{\Delta t=2\times10^{-5}}, compared with the primary \ensuremath{\Delta t=5\times10^{-5}}. Because changing the timestep also changes the Brownian-noise grid, this is a distributional comparison rather than a pathwise or continuum test. We repeated the terminal and fixed-lead structural analyses after either reducing the timestep or replacing the fractional terminal window with a fixed final window of 0.50 time units. Numerical checks and paired-comparison procedures are described in Supplementary Information.

The larger-system comparison crossed 20 \ensuremath{N=128} parents generated independently of the \ensuremath{N=64} primary ensemble with \ensuremath{T_{\rm low}=0.35}, 0.50, or 0.65 and the same four durations. These trajectories used the fine timestep, so the comparison probes persistence of structural outcome discrimination under a compound change in system size and time discretization. Binary AUROC was averaged over durations containing both classes, with parents as bootstrap clusters. Duration-adjusted correlations with terminal mean \ensuremath{\Psi_6} supplied a continuous endpoint when the binary label saturated.

\begin{table*}[!t]
  \centering
  \caption{Simulation and evaluation design. The primary ensemble also supplies the primary-timestep reference for the timestep comparison}
  \label{tab:design}
  \renewcommand{\arraystretch}{1.12}
  \setlength{\tabcolsep}{2.5pt}
  \begin{tabular}{@{}>{\raggedright\arraybackslash}p{0.20\textwidth}c>{\raggedright\arraybackslash}p{0.15\textwidth}>{\raggedright\arraybackslash}p{0.11\textwidth}>{\raggedright\arraybackslash}p{0.13\textwidth}>{\centering\arraybackslash}p{0.08\textwidth}>{\centering\arraybackslash}p{0.07\textwidth}>{\centering\arraybackslash}p{0.11\textwidth}@{}}
    \toprule
    Role & N & Temperature(s) & \ensuremath{\tau_q} & Parent blocks & \shortstack{Replicas\\per set} & Runs & \ensuremath{\Delta t} \\
    \midrule
    Primary ensemble & 64 & \ensuremath{T_{\mathrm{low}}}=0.50 & 0.15, 0.60, 1.50, 3.00 & 40 & 3 & 480 & \ensuremath{5\times10^{-5}} \\
    Fine-step ensemble & 64 & \ensuremath{T_{\mathrm{low}}}=0.50 & same four & 40 & 3 & 480 & \ensuremath{2\times10^{-5}} \\
    Independent larger-system ensemble & 128 & \ensuremath{T_{\mathrm{low}}}=0.35, 0.50, 0.65 & same four & 20 shared across 12 conditions & 1 & 240 & \ensuremath{2\times10^{-5}} \\
    \bottomrule
  \end{tabular}
\end{table*}

\section{Results}

\subsection{Independent noise produces branching from a shared microscopic state}

The three replicas in each set inherited exactly the same pre-quench configuration. Their initial coordinates were identical, so all subsequent separation originated from the independent Brownian histories applied after branching.

\begin{figure*}[tbp]
  \centering
  \includegraphics[width=\textwidth]{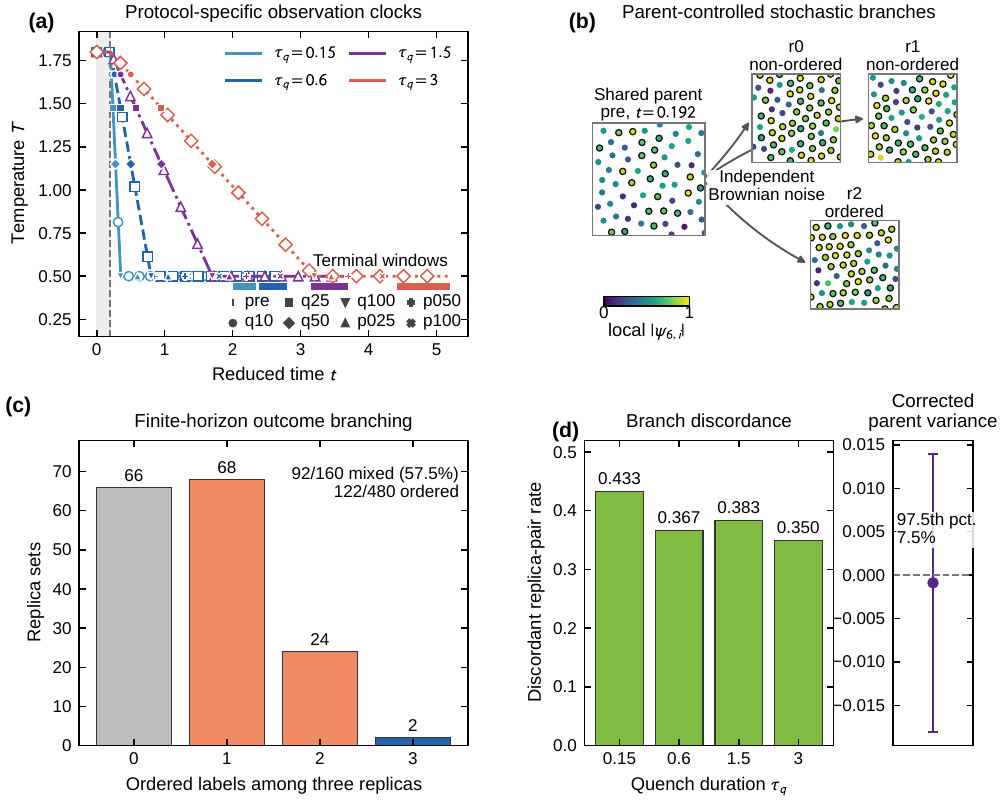}
  \caption{Stochastic branching under a shared initial configuration and quench protocol. (a) Four temperature protocols and the eight sampled phases. The protocols differ only in quench duration; each reaches \ensuremath{T_{\rm low}=0.50} and then remains there for 2.0 reduced time units. Coloured horizontal segments mark the terminal windows. (b) Three independent-noise descendants of one mixed replica set, shown one time unit after quench completion and labelled by their later outcomes. Particle colour denotes local \ensuremath{|\psi_{6,i}|}. (c) Ordered-outcome counts among the three replicas in all 160 sets; 92 sets contain both outcomes. (d) Left: discordant-pair fractions for the four quench durations. Right: decomposition of the observed parent-level spread into finite-replica sampling and the corrected parent component. The error bar is the 95\% complete-parent bootstrap interval, and the dashed line marks zero excess parent variance. Times are in reduced units}
  \label{fig:branching}
\end{figure*}

The 480 trajectories contained 122 ordered terminal outcomes, but these outcomes were distributed broadly within the shared-parent sets. Of the 160 three-replica sets, 66 contained no ordered outcome, 68 contained one, 24 contained two, and 2 contained three. Thus 92 sets (57.5\%) contained both outcomes, and 184 of the 480 within-set replica pairs (38.3\%) ended with different labels (Fig.~\ref{fig:branching}). The observed count distribution was close to the duration-wise Bernoulli expectation (Supplementary Information, Fig. S1).

Equation~\eqref{eq:branch-decomposition} formalizes this branching as a competition between future noise and parent-dependent outcome propensity. After correcting the observed spread for the finite-three-replica contribution, the duration-averaged between-parent component was \ensuremath{-}0.0009 [\ensuremath{-}0.0180, 0.0139]. Its non-negative point estimate was zero, and the upper bootstrap percentile of the parent-attributable fraction was 7.5\%. The three-replica design cannot resolve small differences among individual parent probabilities, but it provides no evidence that such differences dominate the observed branching. More directly, the 92 mixed sets show that a common microscopic parent and protocol did not uniquely determine the later structure.

\subsection{Rapid microscopic decorrelation precedes outcome-associated collective order}

\begin{figure*}[tbp]
  \centering
  \includegraphics[width=\textwidth]{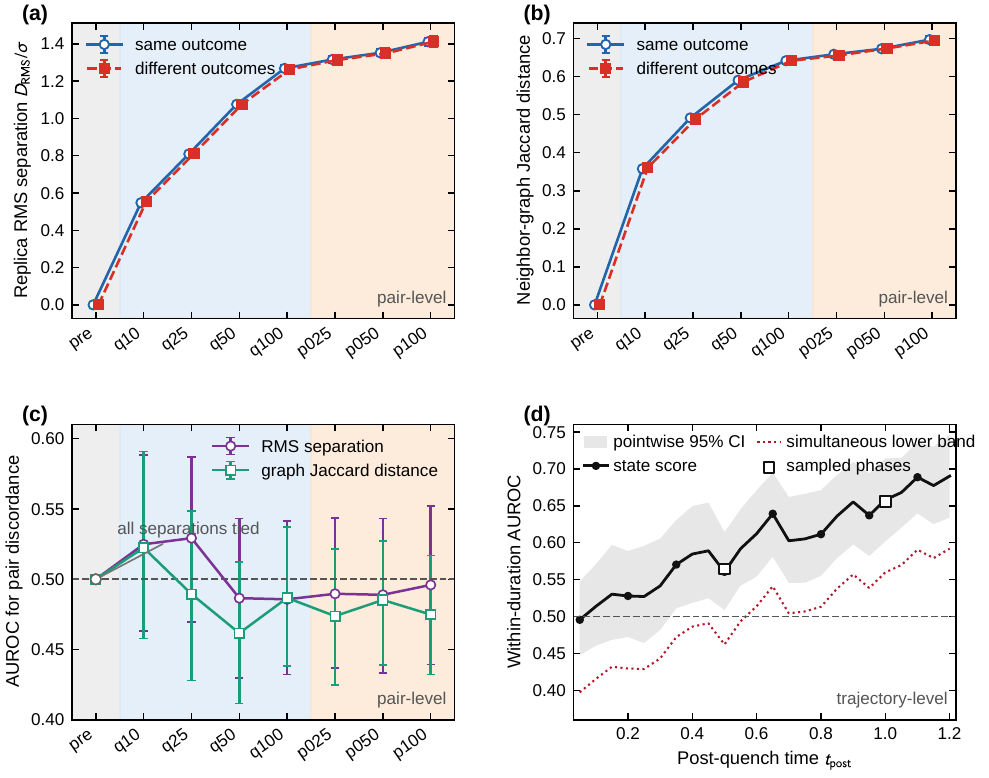}
  \caption{Rapid microscopic decorrelation and the later development of structural outcome discrimination. (a,b) Translation-corrected RMS separation and neighbor-graph Jaccard distance for pairs with the same or different outcomes. Grey, blue, and orange backgrounds denote the shared pre-branch state, quench-relative phases, and post-quench phases. (c) AUROC for distinguishing the two pair types from the separation magnitude; all intervals include 0.5. (d) Duration-averaged state-score AUROC after quench completion. The grey region is the pointwise 95\% interval, the dotted curve is the simultaneous lower band over all 45 sampled times, and open squares show the estimates at \ensuremath{t_{\rm post}=0.50} and 1.00 from the grouped-fold analysis. This lower band remains above 0.5 from \ensuremath{t_{\rm post}=0.60} onward. Times are in reduced units}
  \label{fig:divergence}
\end{figure*}

Independent-noise replicas decorrelated rapidly at the microscopic level. At q10, whose actual elapsed time after branching ranged from 0.008 to 0.296 across the four protocols, the duration-averaged particle-identity RMS separation was approximately 0.55 and the neighbor-graph Jaccard distance approximately 0.36; both continued to grow at later phases (Fig.~\ref{fig:divergence}(a,b)). The same curves were obtained, within uncertainty, for pairs that later shared an outcome and pairs that did not.

The free-particle reference based on time-integrated diffusivity in Eq.~\eqref{eq:free-separation} accounted for the scale and protocol dependence of this separation without making it outcome specific. Across q10, q25, q50, and q100, the measured mean-squared separation had a descriptive through-origin slope $\chi=0.1718$ [0.1669, 0.1769] relative to the free-particle prediction (Supplementary Information, Fig. S16). The observed mean-squared separation was therefore about one-sixth of the free reference across these checkpoints. More importantly, the slopes were 0.1719 [0.1659, 0.1783] for same-outcome pairs and 0.1717 [0.1652, 0.1783] for different-outcome pairs; their difference was \ensuremath{-}0.0002 [\ensuremath{-}0.0087, 0.0085]. The response of the distance to time-integrated diffusivity was thus common to the two pair types.

\begin{figure*}[tbp]
  \centering
  \includegraphics[width=\textwidth]{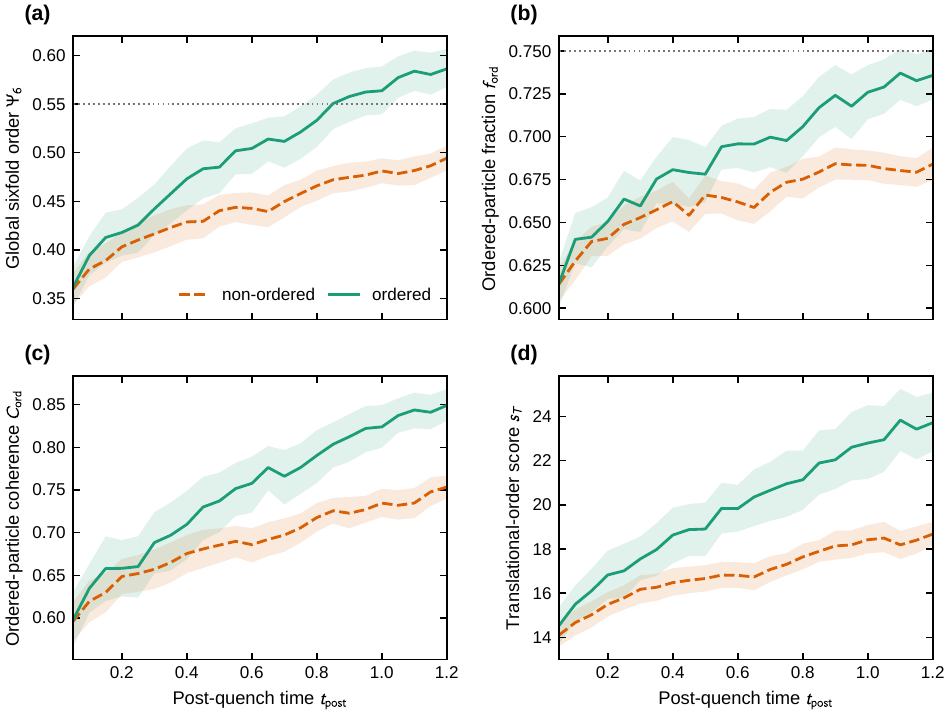}
  \caption{Outcome-resolved development of collective structure after quench completion. Duration-equal means are shown for trajectories with ordered and non-ordered terminal outcomes: (a) global sixfold order \ensuremath{\Psi_6}, (b) ordered-particle fraction \ensuremath{f_{\mathrm{ord}}}, (c) ordered-particle coherence \ensuremath{C_{\mathrm{ord}}}, and (d) reciprocal-space translational-order score \ensuremath{s_T}. Shading gives pointwise 95\% complete-parent bootstrap intervals. Dotted lines in (a,b) mark the corresponding instantaneous structural thresholds; the terminal outcome additionally uses a finite window and a fluctuation criterion. Times are in reduced units}
  \label{fig:pathway}
\end{figure*}

The collective coordinates behaved differently. Immediately after quench completion, the outcome-resolved means were nearly coincident (Fig.~\ref{fig:pathway}). By \ensuremath{t_{\rm post}=0.55}, ordered and non-ordered outcomes had mean \ensuremath{\Psi_6} values 0.502 and 0.444, ordered-particle fractions 0.694 and 0.665, coherences \ensuremath{C_{\mathrm{ord}}} of 0.752 and 0.690, and translational scores \ensuremath{s_T} of 19.84 and 16.82. The gaps continued to grow through \ensuremath{t_{\rm post}=1.20}. These raw structural curves show that the AUROC rise reflects a progressive shift of the physical order distributions rather than only a change in rank tails.

The dense-profile state-score AUROC rose from 0.560 at \ensuremath{t_{\rm post}=0.50} to 0.658 [0.602, 0.714] at 1.00 (Fig.~\ref{fig:divergence}(d)). At \ensuremath{t_{\rm post}=0.55}, raw \ensuremath{\Psi_6}, the state score, \ensuremath{C_{\mathrm{ord}}}, and \ensuremath{s_T} had AUROCs 0.605, 0.592, 0.599, and 0.631, respectively. The separately defined \ensuremath{s_T} had duration-adjusted rank correlations of 0.838--0.872 with the three orientational measures, consistent with coupled translational and orientational development (Supplementary Information, Table S7).

Fixed-lead sampling quantified how far this structural association extended before terminal assessment. At a lead of 1.00, raw \ensuremath{\Psi_6}, the state score, \ensuremath{C_{\mathrm{ord}}}, and \ensuremath{s_T} had equal-duration AUROCs of 0.599, 0.593, 0.605, and 0.593; at a lead of 0.50, these values increased to 0.660, 0.647, 0.646, and 0.673. Changing the terminal window or reducing the timestep preserved simultaneous lower limits above 0.5 for all four measures at leads 0.75 and 0.50, but not at lead 1.00, under the common $4\times4$ orientation-plus-translation family. Because the fixed-window margins at 0.75 were small, we use 0.50 as a conservative common reference for the cross-specification comparison. Complete intervals and duration-resolved results are given in Supplementary Information, Table S6 and Fig. S12.

A scalar potential-energy score and alternative definitions of local and translational order showed the same qualitative development, although their individual time-grid summaries varied (Supplementary Information, Figs. S9, S14, and S15). The evidence therefore supports collective outcome-associated ordering more directly than a unique chronology among its orientational, translational, connectivity, and energetic components.

\subsection{Transient crossings separate early alignment from sustained ordered-particle growth}

\begin{figure*}[!tbp]
  \centering
  \includegraphics[width=\textwidth]{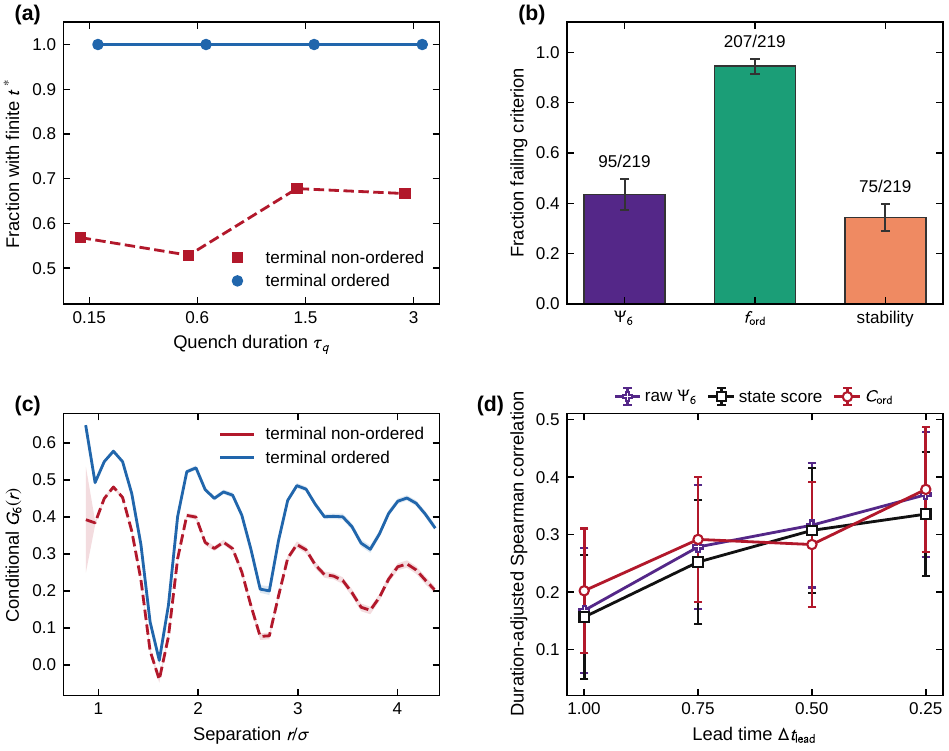}
  \caption{Transient threshold crossings and sustained terminal order. (a) Fraction of trajectories with a finite short-window crossing time \ensuremath{t^*}. All ordered trajectories and 219 non-ordered trajectories cross. (b) Overlapping terminal-criterion failures among those 219 transient crossings; most end below the ordered-particle population criterion. (c) Outcome-resolved bond-orientational pair correlation \ensuremath{G_6(r)}. (d) Duration-adjusted associations of current raw \ensuremath{\Psi_6}, state score, and \ensuremath{C_{\mathrm{ord}}} with the terminal fraction of six-coordinated particles. Reading left to right moves closer to the terminal window as the lead decreases from 1.00 to 0.25. Error bars in (a--c) are pointwise parent-cluster bootstrap intervals; those in (d) are 95\% intervals jointly adjusted across three measures and four leads. Times and leads are in reduced units}
  \label{fig:outcome-structure}
\end{figure*}

The binary terminal labels mark two regions of an underlying continuous structural distribution. The terminal fraction of six-coordinated particles was 0.684 [0.679, 0.690] for ordered outcomes and 0.610 [0.605, 0.615] for non-ordered outcomes. Among the 39 parents that contributed both outcomes, the mean within-parent contrast in this fraction was 0.076 [0.069, 0.083] (Supplementary Information, Fig. S13). The bond-orientational pair correlation \ensuremath{G_6(r)} was likewise higher for ordered outcomes over several neighbor shells, whereas differences in \ensuremath{g(r)} were modest (Fig.~\ref{fig:outcome-structure}(c); Supplementary Information, Fig. S8). These observations place the two outcome classes at different degrees of sixfold order and orientational coherence.

Short-window threshold crossings were frequently reversible. The first-crossing time \ensuremath{t^*} was finite for all 122 terminal ordered trajectories and for 219 of the 358 terminal non-ordered trajectories. Within each quench duration, 52.9--67.8\% of the non-ordered trajectories crossed the short-window rule before returning to the non-ordered terminal class (Fig.~\ref{fig:outcome-structure}(a)). Their first-crossing-time distributions substantially overlapped those of the ordered trajectories (Supplementary Information, Fig. S8).

The subsequent loss of an ordered-particle population was the dominant feature of these transient passages. Of the 219 non-ordered trajectories that crossed, 207 (94.5\% [91.6\%, 97.2\%]) ended below the terminal \ensuremath{f_{\rm ord}} criterion. By comparison, 95 (43.4\%) ended below the \ensuremath{\Psi_6} criterion and 75 (34.2\%) exceeded the fluctuation cutoff; these categories overlap (Fig.~\ref{fig:outcome-structure}(b)). The dense \ensuremath{f_{\rm ord}} profile became persistently outcome-associated later than raw \ensuremath{\Psi_6}, \ensuremath{C_{\mathrm{ord}}}, and \ensuremath{s_T}. Because \ensuremath{f_{\rm ord}} also enters the terminal label, this difference is descriptive rather than a mechanistic onset sequence. It is consistent with a large ordered-particle population being sustained after orientational and translational order have begun to develop.

The association also extended to the terminal fraction of six-coordinated particles, a continuous endpoint that is not part of the binary-label formula. At the longest tested lead, current raw \ensuremath{\Psi_6}, the state score, and \ensuremath{C_{\mathrm{ord}}} had duration-adjusted Spearman correlations of 0.168, 0.156, and 0.202 with this endpoint. Their joint intervals were [0.060, 0.276], [0.048, 0.265], and [0.094, 0.310], respectively (Fig.~\ref{fig:outcome-structure}(d)). The same physical ordering process is therefore visible without reducing the terminal structure to two classes.

\subsection{Protocol, timestep, and larger-system comparisons}

Stochastic branching occurred at all four quench durations, with no detectable monotonic change in ordered-outcome probability over the tested protocol timescales (Supplementary Information, Fig. S1). The aggregate separation between microscopic divergence and collective ordering therefore did not simply track a monotonic outcome-frequency trend, although duration-resolved discrimination was most consistently resolved for the two shorter quenches.

The fine-step ensemble retained the same qualitative growth of post-quench outcome discrimination. Although the sampled time at which a simultaneous lower band cleared 0.5 changed, all four structural measures remained informative at a lead of 0.50 (Supplementary Information, Table S6 and Figs. S7 and S15). The physical conclusion therefore persisted without assigning an exact continuum onset.

\begin{figure*}[tbp]
  \centering
  \includegraphics[width=\textwidth]{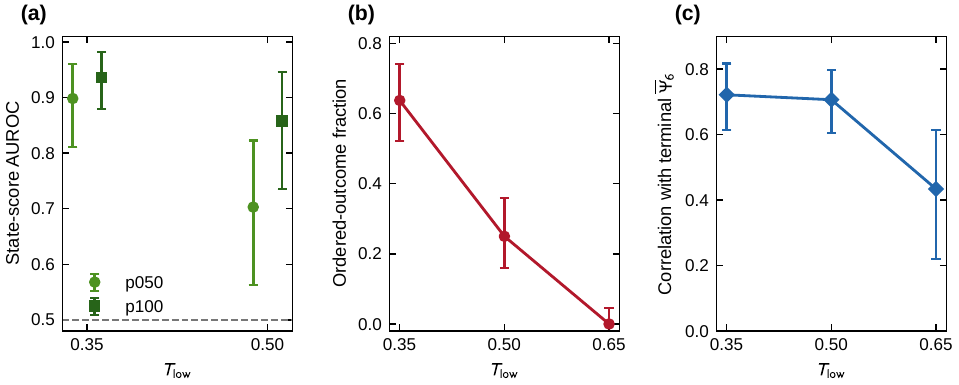}
  \caption{Structural outcome relation for independently generated \ensuremath{N=128} parents. (a) State-score AUROC at \ensuremath{t_{\rm post}=0.50} and 1.00 for temperatures containing both terminal outcomes. (b) Ordered-outcome fraction across the three low temperatures; error bars are descriptive trajectory-level Clopper--Pearson intervals under a Bernoulli reference model, and no inferential comparison relies on them. (c) Duration-adjusted Spearman correlation between the state score at \ensuremath{t_{\rm post}=1.00} and terminal mean \ensuremath{\Psi_6}. Error bars in (a,c) are 95\% complete-parent bootstrap intervals}
  \label{fig:external}
\end{figure*}

The same structural state score showed strong outcome discrimination for independently generated \ensuremath{N=128} parents. At \ensuremath{T_{\mathrm{low}}}=0.35, its AUROCs were 0.898 [0.811, 0.960] at \ensuremath{t_{\rm post}=0.50} and 0.936 [0.880, 0.982] at 1.00. At \ensuremath{T_{\mathrm{low}}}=0.50, the corresponding values were 0.703 [0.563, 0.823] and 0.858 [0.735, 0.946] (Fig.~\ref{fig:external}(a)).

The ordered-outcome fraction decreased from 51/80 at \ensuremath{T_{\mathrm{low}}}=0.35 to 20/80 at 0.50 and 0/80 at 0.65 (Fig.~\ref{fig:external}(b)). As a continuous complement, the state score at \ensuremath{t_{\rm post}=1.00} correlated with terminal mean \ensuremath{\Psi_6} at all three temperatures, with duration-adjusted correlations 0.722, 0.707, and 0.434, respectively (Fig.~\ref{fig:external}(c)). This relation therefore remained measurable at \ensuremath{T_{\mathrm{low}}}=0.65, where the binary label saturated and AUROC was undefined.

Graph models recovered the late outcome association but did not reveal a comparably supported earlier or independent coordinate. Adding the tested graph scores to state and duration produced no detectable held-out gain, and feature sensitivities pointed back to local sixfold order (Supplementary Information, Figs. S3, S4, and S11). They therefore serve as a representation control rather than the physical basis of the conclusion.

\section{Discussion}

The shared-parent ensemble separates two processes that are often conflated. Equation~\eqref{eq:branch-decomposition} shows how a common parent can still have a distribution of future outcomes, and the mixed replica sets demonstrate that this stochastic branching is frequent in the present ensemble. Equation~\eqref{eq:free-separation} then provides the natural thermal scale for the initial loss of microscopic similarity. Interparticle forces reduce the separation relative to the free reference, but the same exposure response is observed for pairs that later share an outcome and pairs that do not. Generic decorrelation is therefore not itself an indicator of assembly fate.

This result has a simple high-dimensional interpretation. Coordinate RMS and graph Jaccard distances collect displacements or edge changes across all available modes. Most of those changes need not point toward or away from collective order, so an unweighted norm can grow while outcome-relevant information remains diluted. Sixfold alignment, ordered-particle coherence, and first-shell translational order instead project the evolving configuration onto symmetry-sensitive collective modes. Their outcome-resolved curves show a gradual physical separation after the quench, rather than only a classifier-score effect. Intermediate-time reshooting would be required to test whether any of these coordinates approximates a finite-time committor \citep{Bolhuis2002TPS,E2006TransitionPaths,Chen2023ReactionPathways,ContrerasArredondo2026Committor}.

The later dynamics also distinguish the appearance of order from its persistence. The translational score \ensuremath{s_T} uses neither bond angles nor the terminal-label thresholds, yet evolves with the orientational coordinates. Current structure is associated with the terminal fraction of six-coordinated particles, which is not included as a separate term in the binary-label formula. Moreover, short-window threshold crossings occur in both outcome classes, while 94.5\% of transient non-ordered crossings subsequently lose the required ordered-particle population. Successful assembly in this finite observation window is therefore associated with maintaining an extended ordered population, not simply reaching a high instantaneous alignment once.

The fixed-lead analysis places this developing association on a common distance from terminal assessment. A lead of 0.50 corresponds to \ensuremath{0.25\tau_B} at \ensuremath{T_{\mathrm{low}}}, although this free-particle conversion is not an interacting relaxation time. All four structural measures remained informative at this lead after either changing the terminal window or reducing the timestep; state-score discrimination also persisted in an independently generated \ensuremath{N=128} ensemble. The available data do not select a universal clock for the collective ordering process, so the result is a finite-protocol dynamical separation rather than a universal onset law. The graph-model controls reach the same late structural information without revealing an independent earlier coordinate.

\subsection{Scope and limitations}
The terminal labels describe degrees of finite-window order during a two-unit post-quench hold in an incommensurate square cell. They do not assign equilibrium phases or establish irreversible nucleation or asymptotic kinetic arrest. Three replicas per parent also limit the resolution of parent-specific outcome probabilities; the 7.5\% value is an upper bootstrap percentile of a variance fraction, not a predictive-accuracy bound. The free-draining dynamics omit hydrodynamic interactions.

The timestep comparison changes both the timestep and Brownian-noise grid, and the \ensuremath{N=128} comparison changes size and time discretization together. Matched timestep and size hierarchies would be required for continuum or finite-size extrapolation. Independent measurements of the structural relaxation time \ensuremath{\tau_\alpha} would additionally express the protocol durations and leads in relaxation-time units. Because orientational and translational order are accessible in particle-resolved imaging, the separation between rapid decorrelation and developing collective order can also be tested in colloidal experiments \citep{Gasser2001RealSpaceNucleation,Zahn1999TwoStage,Lavergne2017GrainGrowth}.

\section{Conclusion}

Independent Brownian histories rapidly decorrelate replicas launched from a shared microscopic state. Their mean-squared separation remains below the free-particle reference based on time-integrated diffusivity, but the fitted response is the same for replica pairs with equal and different terminal outcomes. The magnitude of microscopic separation therefore carries little information about the later assembly fate.

Outcome association develops instead in collective structure after quench completion. Sixfold orientational order, ordered-particle population and coherence, and translational order progressively separate the terminal outcomes. A fixed-lead comparison spanning global order, the state score combining global and population order, coherence, and translational order remains informative at a tested lead of 0.50 after changing either the terminal window or the integration timestep. Transient crossings show that instantaneous alignment is insufficient: most non-ordered crossings subsequently lose the extended ordered-particle population. Quenched assembly in this ensemble thus combines generic noise-driven decorrelation with the later emergence and finite-window persistence of collective order.

\backmatter

\medskip\noindent\textbf{Supplementary information}\par

The Supplementary Information follows the main article and contains extended methods, additional analyses, 16 supplementary figures, and seven supplementary tables.

\medskip\noindent\textbf{Statements and declarations}\par

\smallskip\noindent\textbf{Competing interests}\par

The authors declare no competing interests.

\smallskip\noindent\textbf{Data and code availability}\par

The original trajectories, processed trajectory-level data, out-of-fold predictions, bootstrap summaries, Brownian-dynamics and observable implementations, and graph-model training source are archived on Zenodo at \url{https://doi.org/10.5281/zenodo.22179833}. The archive includes relative-path manifests, fixed training seeds, and scripts for reproducing the principal results.

\bibliography{references}

\end{document}


\maketitle

This Supplementary Information gives the derivations and extended methods that support
the physical comparison between microscopic replica separation and collective
ordering. It also documents the outcome definition, grouped statistical
analysis, graph-model representation control, timestep and terminal-window
comparisons, and additional structural observables. All analyses use the
trajectories described below; model-based analyses additionally use held-out
predictions.

\section{Supplementary methods}

\subsection{Quench-duration analysis}

For each quench duration $\tau_q\in\{0.15,0.60,1.50,3.00\}$, the primary
ensemble contained 40 parent states and three noise replicas per parent.
Descriptive trajectory-level Clopper--Pearson intervals under a Bernoulli
reference model accompany the duration-specific ordered-label fractions; no
inferential comparison relies on these intervals. Mixed
replica sets contain both terminal outcomes among the three replicas, and pair
discordance is the fraction of the three within-set replica pairs with
different labels.  The duration contrast was specified on $\log\tau_q$ and
estimated with parent-cluster uncertainty; its slope was $-0.0135$ with 95\%
interval $[-0.0465,0.0200]$.

\begin{figure}[htbp]
\centering
\includegraphics[width=\textwidth]{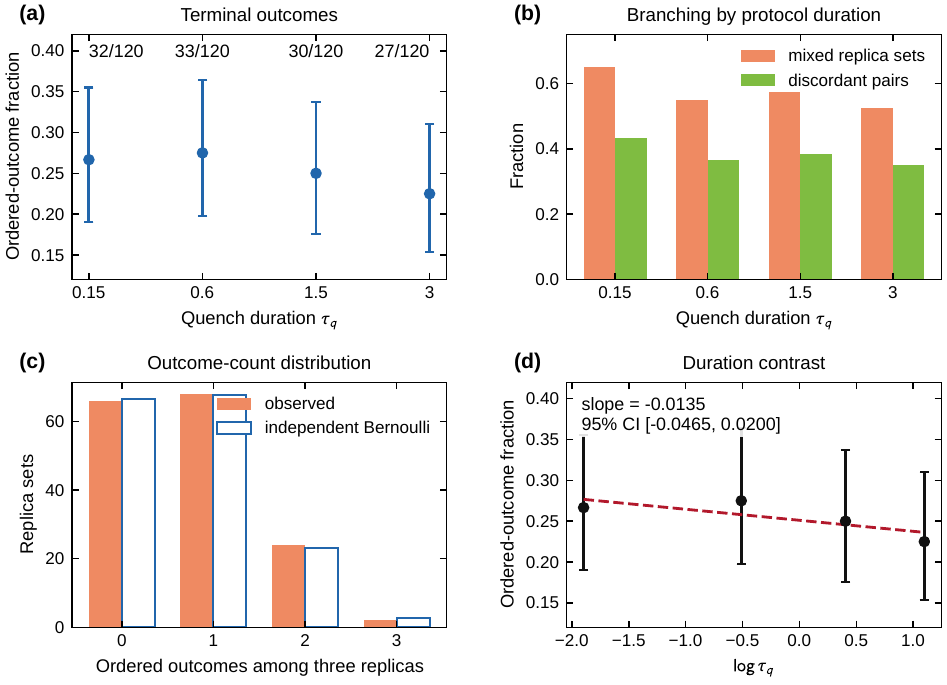}
\caption{\textbf{Quench-duration design and terminal outcome frequencies.}
(a) Ordered-label fraction and descriptive trajectory-level Clopper--Pearson
interval under a Bernoulli reference model by $\tau_q$; no inferential
comparison relies on these intervals. (b) Mixed-set
and discordant-pair fractions. (c) Observed distribution of zero to three
ordered labels per replica set compared with independent Bernoulli expectations formed
within duration; the expected mixed-set count is 90.8, compared with 92
observed. (d) Log-duration contrast. The slope is
$-0.0135$ $[-0.0465,0.0200]$, and its interval includes zero
}
\label{fig:s1}
\end{figure}

The corresponding duration-level outcomes and replica-set counts are shown in
Fig.~\ref{fig:s1}; the complete ensemble design is summarized in
Table~\ref{tab:s1design}.

\subsection{Relative diffusion and finite-time branching}

Let $t_b=t_{\rm hold}$ denote the instant at which replicas begin to use
independent Brownian noise. Subtracting the overdamped Langevin equations for
replicas $a$ and $b$ gives
\begin{equation}
 d\delta\mathbf r_i
 =\mu\left(\mathbf F_i^{(a)}-\mathbf F_i^{(b)}\right)dt
 +\sqrt{4D(t)}\,d\mathbf W_i,
 \qquad D(t)=\mu k_{\rm B}T(t).
 \label{eq:s-relative-motion}
\end{equation}
For two non-interacting replicas in two dimensions, each Cartesian relative
coordinate has variance $4\int_{t_b}^{t}D(s)\,ds$. Removing the common
translation subtracts the sample mean of the $N$ displacement vectors and
multiplies the particle-averaged variance by $1-1/N$. Hence
\begin{equation}
 X(t)\equiv\left\langle D_{\rm RMS}^{2}(t)\right\rangle_{\rm free}
 =8\left(1-\frac{1}{N}\right)\int_{t_b}^{t}D(s)\,ds.
 \label{eq:s-free-separation}
\end{equation}
For the linear temperature ramp, at fractional quench coordinate
$\alpha=(t-t_b)/\tau_q\in[0,1]$,
\begin{equation}
 X(\alpha,\tau_q)=8\left(1-\frac{1}{N}\right)\mu k_{\rm B}\tau_q
 \left[T_{\rm high}\alpha-
 \frac{T_{\rm high}-T_{\rm low}}{2}\alpha^2\right].
 \label{eq:s-exposure-quench}
\end{equation}
After quench completion this becomes
\begin{equation}
 X(t_{\rm post},\tau_q)=8\left(1-\frac{1}{N}\right)\mu k_{\rm B}
 \left[\frac{T_{\rm high}+T_{\rm low}}{2}\tau_q
 +T_{\rm low}t_{\rm post}\right].
 \label{eq:s-exposure-post}
\end{equation}
Equations~\eqref{eq:s-free-separation}--\eqref{eq:s-exposure-post} define a
free-particle reference, not an effective theory for the interacting liquid.
We fitted the 16 duration-by-checkpoint combinations at q10, q25, q50, and q100
to $\langle D_{\rm RMS}^{2}\rangle=\chi X$ through the origin. Each
combination averages the available replica pairs. Complete-parent
resampling preserved all replica pairs and quench durations. The descriptive
slope was $\chi=0.1718$ $[0.1669,0.1769]$ for all pairs, $0.1719$
$[0.1659,0.1783]$ for same-outcome pairs, and $0.1717$
$[0.1652,0.1783]$ for different-outcome pairs. Their difference was
$-0.0002$ $[-0.0087,0.0085]$ (Fig.~\ref{fig:sexposure}). The force-difference term
in Eq.~\eqref{eq:s-relative-motion} makes $\chi$ history dependent, so these
four checkpoints do not establish a universal rescaled clock.

The replica design also yields an exact decomposition of finite-time outcome
variation. Let $Y_{p,c}^{(a)}\in\{0,1\}$ be the outcome of replica $a$ from
parent $p$ under protocol $c$, and let
$q_{p,c}=P(Y_{p,c}^{(a)}=1\mid p,c)$. Conditional independence of two future
noise realizations gives
\begin{equation}
 P\!\left(Y_{p,c}^{(a)}\ne Y_{p,c}^{(b)}\mid p,c\right)
 =2q_{p,c}(1-q_{p,c}).
\end{equation}
Writing $p_c=E_p(q_{p,c})$ and averaging the discordance probability over
parents as $d_c$ gives
\begin{align}
 \frac{d_c}{2}&=E_p(q_{p,c})-E_p(q_{p,c}^{2}),\\
 p_c(1-p_c)&=\frac{d_c}{2}+\operatorname{Var}_p(q_{p,c}).
 \label{eq:s-branch-decomposition}
\end{align}
Thus the Bernoulli variation at fixed protocol separates into within-parent
future-noise branching and between-parent variation in finite-time outcome
propensity. With three replicas, if $k_{p,c}$ are ordered, an unbiased estimate
of $q_{p,c}(1-q_{p,c})$ is
\begin{equation}
 \widehat W_{p,c}=\frac{k_{p,c}(3-k_{p,c})}{3(3-1)}.
 \label{eq:s-within-estimator}
\end{equation}
The finite-replica correction and parent-attributable fraction are specified
below with the dense-time uncertainty analysis.

\subsection{Grouped model evaluation}

The baseline and duration-conditioned graph neural networks (GNNs) were evaluated at eight predefined
phases. The targeted GNN sensitivity model was evaluated at q100, p025, p050,
and p100. Each model--phase combination used five parent-grouped outer folds
and three repeated splits. All descendants of one parent remained in the same
fold. Within every fold--duration cell there were 24 held-out trajectories, and
every original cell contained both outcome classes. An out-of-fold prediction
was therefore obtained for each of the 480 trajectories in every repeat.
Within-duration AUROCs were given equal weight, calculated separately for each
repeat, and then averaged. A cell that contained one class after parent
resampling was omitted only within that bootstrap draw.

Checkpoint selection remained internal to the training and validation portion
of each outer fold. Baseline and duration-conditioned runs maximized a
duration-balanced validation AUROC; targeted-GNN runs maximized the mean of the
defined duration-specific validation AUROCs. Because the targeted model also
changed its input channel, history sampling, and auxiliary objective, it is a
combined sensitivity analysis rather than a single-feature ablation.

For deterministic state-score checkpoint comparisons, the same fold
assignments preserved the evaluation domain. Its repeat variation reflects
only the fold partition, whereas the learned-model repeat layer also includes
training variability. Dense structural profiles do not use fold partitions.

\subsection{Observation histories and window separation}

At every requested cutoff, analysis used the latest stored frame not later than
that time. The baseline and duration-conditioned GNNs sampled eight
frames from each protocol segment available by the cutoff, giving histories of
8, 16, or 24 selected frames. The targeted sensitivity model instead used the
last eight consecutive distinct frames. Repeated frame indices produced zero
displacement in the corresponding local input.

For all 32 phase-by-duration cells, the complete observation history was
strictly disjoint from the terminal label window. The minimum nominal lead to
the terminal-window start was 0.232, and the minimum lead after mapping to the
actual stored frame was 0.240. The dense structural grid comprised 21
quench-relative coordinates, $\alpha=0,0.05,\ldots,1$, evaluated at
$t=t_{\rm hold}+\alpha\tau_q$, and 24 absolute post-quench coordinates,
$s=0.05,0.10,\ldots,1.20$, evaluated at
$t=t_{\rm hold}+\tau_q+s$. At $t_{\rm post}=1.20$, the minimum lead to the
primary fractional terminal window was 0.032 both nominally and after
stored-frame selection; for the fixed final window, the corresponding leads
were 0.300 and 0.304.

\subsection{Graph representations and training}

At each selected frame, particles were nodes of the $r_{\rm nb}=1.5$ radius
graph. The baseline node vector contained six invariant local channels:
coordination divided by 6, mean neighbor distance divided by 1.5,
neighbor-distance standard deviation divided by 1.5, local $|\psi_6|$,
coordination defect $|n_i-6|/6$, and the scalar magnitude of the periodic
minimum-image displacement from the preceding selected frame,
$|\Delta\mathbf r_i|/1.5$. The displacement magnitude was zero for the first
selected frame; a repeated frame index also produced zero displacement.

Three GraphSAGE layers of width 64 generated node embeddings. Mean pooling
produced one frame embedding, and a 64-unit long short-term memory (LSTM) network encoded the selected history.
The baseline objective combined weighted binary cross-entropy with SmoothL1
regression of the terminal-window mean disordered-particle fraction
$1-\langle\ford\rangle_{\rm term}$ and masked regression of
$t^*/t_{\max}$, with weights 1, 0.5, and 0.2. Here $t_{\max}$ is the final
stored time of the corresponding trajectory. The duration-conditioned model
retained this representation and objective but concatenated standardized
$\log\tau_q$ to every pooled frame embedding before the LSTM. Its fixed center
and scale were $-0.225967$ and 1.121061.

The targeted GNN sensitivity model added the local phase-coherence channel $c_{6,i}$, used
the three 64-wide GraphSAGE layers and 64-unit LSTM, used the last eight
consecutive distinct frames, and optimized binary classification alone. It was
evaluated at q100, p025, p050, and p100; p050 and p100 were used for the
state-adjusted residual analysis. Because the input, history, auxiliary losses,
and selection rule change together, the targeted-GNN comparison is not a
single-feature ablation.

AdamW used learning rate $10^{-3}$, weight decay $10^{-4}$, and batch size 32.
Training ran for at most 40 epochs with gradient-norm clipping at 1 and
early-stopping patience 8. Class weights came only from the training fold.
Baseline and duration-conditioned checkpoints maximized duration-balanced
validation AUROC; targeted-GNN checkpoints maximized the macro-average of defined
duration-specific validation AUROCs. Selection remained internal to each outer
fold.

\subsection{Terminal labels and local definitions}

The primary label used the final 15\% of stored frames, whereas the fixed-window
label used the final 0.50 absolute time units. The latter produced 121
ordered labels and 91.46\% agreement with the primary labels. Local-order
sensitivities changed the neighbor cutoff from $r_{\rm nb}=1.5$ to 1.4 or 1.6
while retaining coordination range 5--7. The 5--7 tolerance accommodates
thermal and cutoff-based coordination fluctuations around locally sixfold
environments. Requiring exactly six neighbors retained the original
$\ford\ge0.75$ threshold without recalibration and produced zero ordered labels;
binary AUROC is therefore undefined, not equal to 0.5. This sensitivity tests
the joint definition and threshold rather than a competing calibrated phase
definition.

The primary numerical cutoffs, $\operatorname{mean}(\Psi_6)\ge0.55$,
$\operatorname{mean}(\ford)\ge0.75$, and
$\operatorname{sd}(\Psi_6)\le0.08$ over the final 15\% of stored frames, were
used to define the terminal label. They remained fixed during training and
the subsequent dense-time comparisons and are not interpreted as fitted
thermodynamic phase boundaries.

\begin{figure}[p]
\centering
\includegraphics[width=\textwidth]{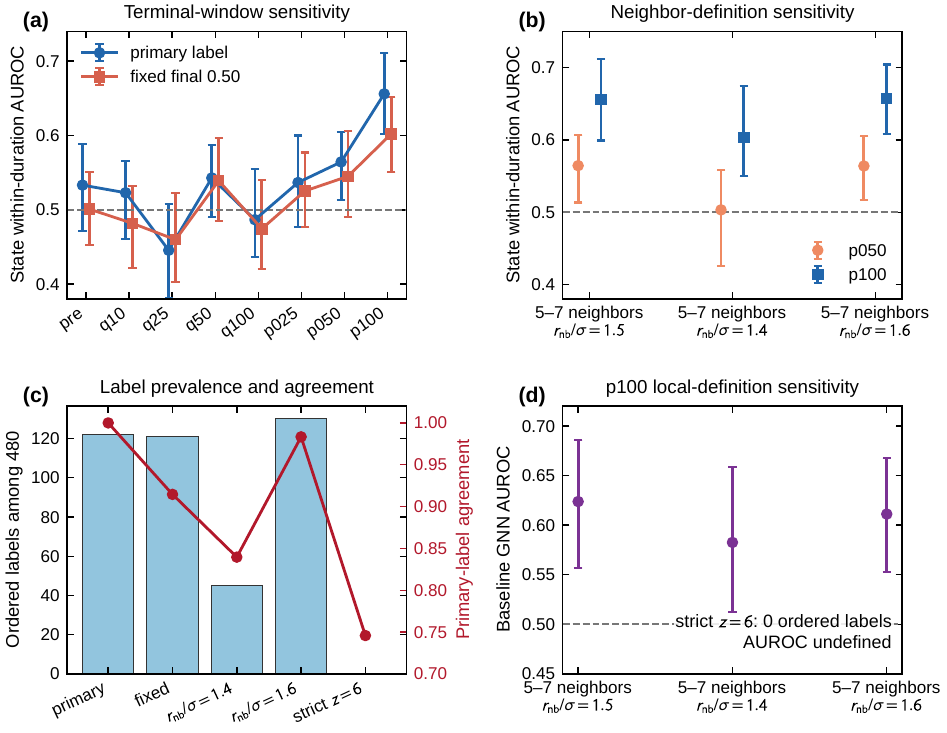}
\caption{\textbf{Terminal-label and neighbor-definition sensitivity.}
(a) Predefined state-score AUROC under the primary and fixed absolute terminal windows.
(b) State p050 and p100 estimates for the primary neighbor definition and
cutoff variants $r_{\rm nb}/\sigma=1.4$ and 1.6, all retaining the 5--7
coordination range.
(c) Label prevalence and agreement with the primary definition. (d) Baseline
GNN p100 sensitivity. The strict $z=6$ definition keeps the original ordered-fraction threshold
without recalibration; it yields no ordered labels and hence has
undefined AUROC}
\label{fig:s3}
\end{figure}

The sensitivity analysis comprised eleven label variants. One-at-a-time variants used
$\Psi_6$ cutoffs 0.50 or 0.60, $\ford$ cutoffs 0.70 or 0.80, stability cutoffs
0.10 or 0.06, and final-window fractions 0.10 or 0.20. Relaxed and strict joint
variants moved all three structural cutoffs together. Only terminal outcomes
were recomputed. All predictor scores were held fixed: stored baseline GNN
predictions were rescored without retraining, and the baseline state score
retained its original denominators 0.55 and 0.75 rather than adopting each
variant's label thresholds. The same above-chance criterion required
the point estimate and both the parent-cluster and hierarchical seed--parent lower limits
to exceed 0.5. Figure~\ref{fig:s3} and Table~\ref{tab:s5labels} report the
label, neighbor-definition, and discrete state/GNN sensitivities.

\subsection{Additional orientational observables}

The main text defines the ordered mask, its population fraction $\ford$, and
the within-mask coherence $\Cord$. We additionally define $C_{\max}$ as the
number of particles in the largest connected component of the
$r_{\rm nb}=1.5$ radius graph induced by the ordered mask; $\Cmax$ was set to
zero when the mask was empty. The remaining orientational observables and model
inputs were defined as follows. To regularize the local phase, we used
\begin{equation}
 \widetilde u_i=\psi_{6,i}/\max(|\psi_{6,i}|,10^{-12}).
\end{equation}
The rotation-invariant local phase-coherence input of the targeted GNN
was
\begin{equation}
 c_{6,i}=n_i^{-1}\sum_{j\in\mathcal N_i}
 \operatorname{Re}(\widetilde u_i\widetilde u_j^*),
\end{equation}
with $c_{6,i}=0$ when $n_i=0$. The denominator remains $n_i$, so a site with
$|\psi_{6,i}|<10^{-12}$ contributes with attenuated magnitude rather than an
assigned unit phase.

Let $E_\epsilon$ contain every undirected $r_{\rm nb}$ edge whose endpoints
both satisfy $|\psi_6|>\epsilon$, with $\epsilon=10^{-12}$. The neighbor-wide
phase coherence was
\begin{equation}
 \Cnb=|E_\epsilon|^{-1}\sum_{(i,j)\in E_\epsilon}
 \operatorname{Re}(\widetilde u_i\widetilde u_j^*),
\end{equation}
and was set to zero for an empty edge set. Unlike $\Cord$, $\Cnb$ is not
restricted to the ordered mask. Endpoint cutoffs 0.1 and 0.3 and an
amplitude-weighted edge average proportional to
$|\psi_{6,i}||\psi_{6,j}|$ supplied definition-level sensitivities.

Because $\Cord$ normalizes within the ordered set, its population-weighted
counterpart was
\begin{equation}
 A_{\rm ord}=N^{-1}\left|\sum_{i\in O}\psi_{6,i}\right|
 =\Cord\,N^{-1}\sum_{i\in O}|\psi_{6,i}|.
\end{equation}
The ordered-set-size distribution and minimum requirements of 3, 5, or 8
ordered particles were checked before this normalization sensitivity was
interpreted. These definitions separately probe population, orientational
alignment, and graph connectivity; their sampled discrimination profiles do
not by themselves establish a mechanistic chronology.

\subsection{Feature dependence and current-state analyses}

Test-time mean replacement substituted one baseline node channel by its
training-fold mean and reported intact minus ablated AUROC.  Integrated
gradients used the trained ordered-label-logit margin; absolute shares sum to unity
within phase.  Both quantities depend on the trained reference model and
correlated input channels and are not interpreted as causal feature effects.

For state-adjusted residual tests, the null logistic stacker contained a cubic
B-spline transform of the predefined state score and categorical quench duration.
The transform used four uniformly spaced knots, degree 3, and no bias column and was
fitted on training parents only. The full stacker added one GNN logit after
clipping its probability to $[10^{-5},1-10^{-5}]$. Both stackers used
L2-regularized logistic regression ($C=1$, \texttt{lbfgs}, at most 2000
iterations), were fitted to out-of-fold (OOF) predictions from the four training
parent folds, and were evaluated on the held-out
parent fold. The threshold-defined subset excluded
trajectories already satisfying both instantaneous thresholds
$\Psi_6\ge0.55$ and $\ford\ge0.75$.  A separate parent-grouped cross-fitted
spline-ridge analysis tested whether current $\Cord$ improved prediction of
terminal-window mean $\Cmax$ after current $\Psi_6$, $\ford$, $\Cmax$, and
$\tau_q$ were included.

\begin{figure}[p]
\centering
\includegraphics[width=\textwidth]{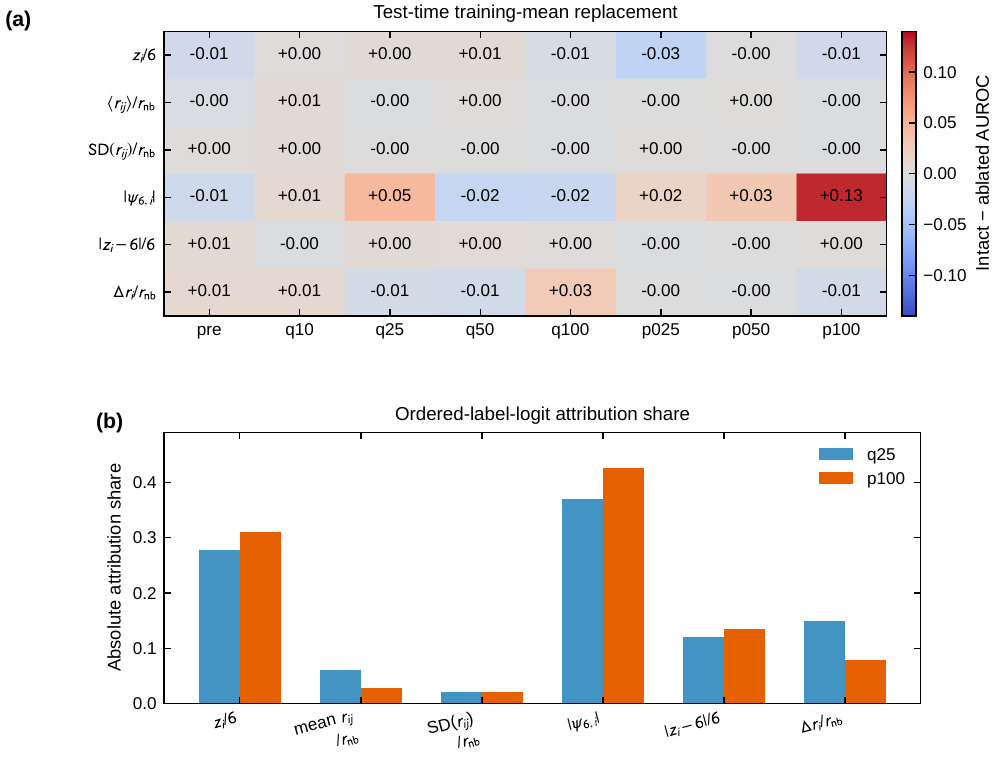}
\caption{\textbf{Baseline-model feature dependence.} (a) Test-time
training-mean replacement across phases and node channels. (b) Absolute
integrated-gradient shares at q25 and p100. The six plotted channels are
$z_i/6$, $\langle r_{ij}\rangle/r_{\rm nb}$,
${\rm SD}(r_{ij})/r_{\rm nb}$, $|\psi_{6,i}|$, $|z_i-6|/6$, and
$\Delta r_i/r_{\rm nb}$. Local $|\psi_6|$ dominates the
late p100 dependence.  The quantities are reference-dependent model
sensitivities rather than causal effects}
\label{fig:s4}
\end{figure}

\begin{figure}[p]
\centering
\includegraphics[width=\textwidth]{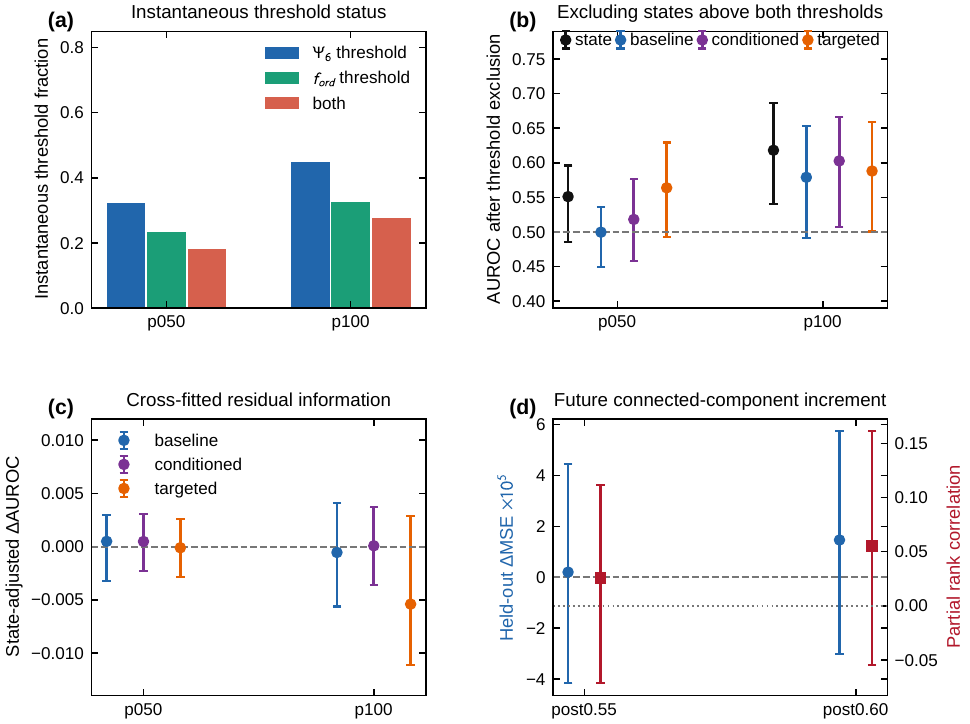}
\caption{\textbf{Current-state and residual-information analyses.} (a)
Instantaneous threshold rates at p050 and p100. (b) AUROC after excluding
trajectories that already meet both structural thresholds; error bars are
parent-cluster bootstrap intervals. The p100 targeted-GNN lower limit is 0.501, but
its hierarchical seed--parent bootstrap interval $[0.481,0.672]$ crosses 0.5. (c) Cross-fitted state-adjusted GNN
increments. (d) Incremental association of current $\Cord$ with later $\Cmax$
after controlling for current state and duration. All intervals for the residual
and coherence increments include zero}
\label{fig:s5}
\end{figure}

Feature-dependence estimates are shown in Fig.~\ref{fig:s4}; analyses of the
current state, the threshold-defined subset, residual information, and future
growth are shown in
Fig.~\ref{fig:s5}.

\subsection{Dense-time bands and finite-replica correction}

The extended grid comprised 21 quench-relative sampled times and 24 post-quench
times through $t_{\rm post}=1.20$, for 45 times and $480\times45=21\,600$
trajectory--time observations. Six observables were compared: raw global
$\Psi_6$, the state score, $\Cord$, $\Cnb$, $\ford$, and $\Cmax$. Let
$\widehat\theta_{jk}$ be the within-duration AUROC for observable $j$ and
sampled time $k$, and let $\widehat\theta^{(b)}_{jk}$ be its value in a parent-cluster
bootstrap draw $b$. Pointwise limits were the 2.5th and 97.5th percentiles
across $b$.

Two correction families were used. The full 45-time family calibrated over all
quench-relative and post-quench times, whereas the 24-time post-quench family
calibrated over only the absolute post-quench times. For time set
$\mathcal K$, the observable-wise radius was
\begin{equation}
 r_j(\mathcal K)=Q_{0.95}\!\left[\max_{k\in\mathcal K}
 \left|\widehat\theta^{(b)}_{jk}-\widehat\theta_{jk}\right|\right],
\end{equation}
and the joint radius was
\begin{equation}
 r_{\rm joint}(\mathcal K)=Q_{0.95}\!\left[\max_j\max_{k\in\mathcal K}
 \left|\widehat\theta^{(b)}_{jk}-\widehat\theta_{jk}\right|\right].
\end{equation}
The simultaneous bands were $\widehat\theta_{jk}\pm r_j$ or
$\widehat\theta_{jk}\pm r_{\rm joint}$. The same parent draw was used for all
observables and times. The earliest persistently discriminative sampled time was the first post-quench
grid point whose lower band remained above 0.5 through
$t_{\rm post}=1.20$. AUROC 0.60 was retained only as a descriptive effect-size
reference. All headline intervals used 5000 complete-parent draws. Increasing
the dense-band calculation to $100\,000$ draws left the reported structural
conclusion unchanged.

For parent $p$ and duration $c$, let $z_{p,c}=k_{p,c}/3$. Using the within-parent
estimator in Eq.~\eqref{eq:s-within-estimator}, the raw duration-specific
between-parent component was
\begin{equation}
  \widehat V_c=\operatorname{var}_p(z_{p,c})-
  \frac{\operatorname{mean}_p(\widehat W_{p,c})}{3}.
\end{equation}
The four raw components were averaged with equal duration weight before one
non-negative projection,
\begin{equation}
  \widehat V=\frac{1}{4}\sum_c\widehat V_c,
  \qquad \widehat V_+=\max(\widehat V,0).
\end{equation}
The parent-attributable fraction was
\begin{equation}
  f_{\rm parent}=\frac{\widehat V_+}
  {\operatorname{mean}_c\operatorname{mean}_p(\widehat W_{p,c})+\widehat V_+}.
\end{equation}
Every bootstrap draw repeated this operation in the same order.  The raw point
was $-0.0009$ with interval $[-0.0180,0.0139]$; the point after projection was
zero and the 97.5th percentile of $f_{\rm parent}$ was 7.5\%.

\begin{figure}[p]
\centering
\includegraphics[width=\textwidth]{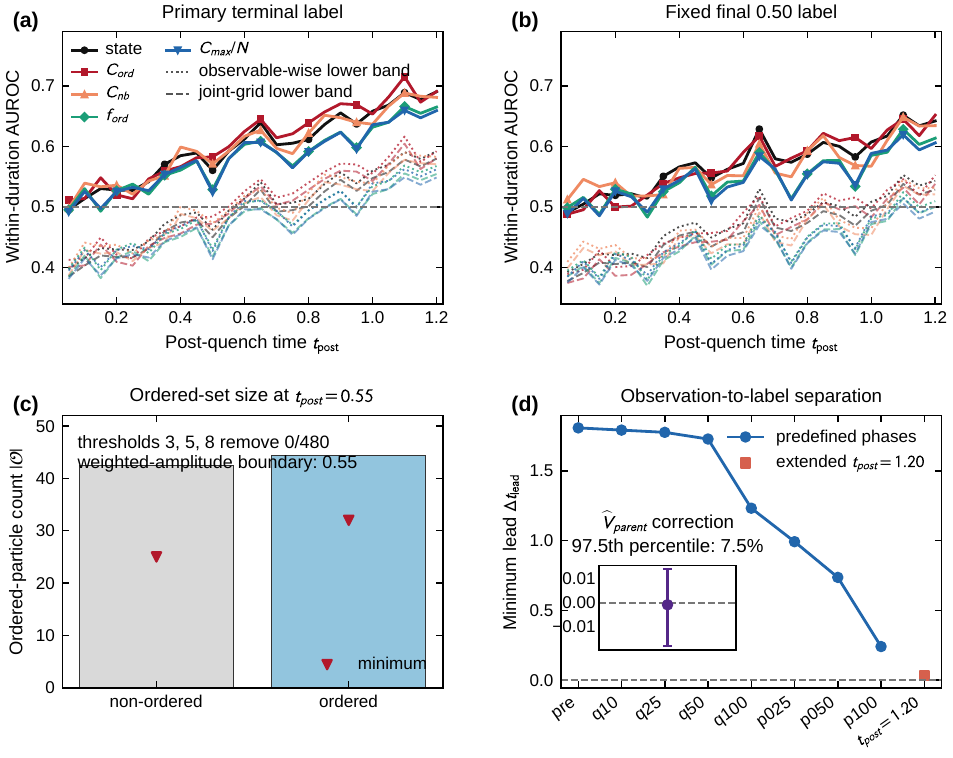}
\caption{\textbf{Dense sampled times, lead time, ordered-set size, and finite-replica
correction.} Post-quench portions of the 45-point curves with observable-wise
(dotted) and joint observable-by-time (dashed) simultaneous lower bands for
(a) primary and (b) fixed-window labels. (c)
Ordered-particle counts at $t_{\rm post}=0.55$; thresholds of 3, 5, and 8
remove no trajectory, and the population-weighted amplitude gives the same
earliest persistently discriminative sampled time of 0.55. (d) Minimum
actual observation-to-primary-label lead for the predefined phases and the
$t_{\rm post}=1.20$ extension, with the corrected parent-variance estimate inset
}
\label{fig:s6}
\end{figure}

Figure~\ref{fig:s6} displays the dense bands, lead-time analysis, ordered-set
counts, and finite-replica correction. The principal outcome-resolved
structural profiles appear in main-text Fig.~3.

\subsection{First-reciprocal-shell translational-order comparison}

The translational comparison used only instantaneous coordinates, number density,
and periodic-box geometry. For every allowed reciprocal vector
$\mathbf{k}=(2\pi/L)(n_x,n_y)$, we calculated
\begin{equation}
 S(\mathbf{k})=\frac{1}{N}\left|\sum_{j=1}^{N}
 \exp(i\mathbf{k}\cdot\mathbf{r}_j)\right|^2.
\end{equation}
The ideal triangular reference at density $\rho$ has
$a_{\triangle}=[2/(\sqrt{3}\rho)]^{1/2}$ and
$k_0=4\pi/(\sqrt{3}a_{\triangle})$. At $\rho=0.82$ and
$L=8.834522$, $a_{\triangle}=1.186664\sigma$, $k_0=6.113944/\sigma$,
and the fundamental reciprocal spacing is $2\pi/L=0.711208/\sigma$.
We retained one vector from each inversion pair in the annulus
$\bigl|\,|\mathbf{k}|-k_0\,\bigr|\leq\pi/L$. This produced 24 vectors on squared integer
shells 68, 72, 73, 74, 80, 81, and 82.

For each frame, the primary score $\sT$ was the mean of the three largest
annular $S(\mathbf{k})$ values. Three was fixed to match the number of
inversion-unique directions in an ideal triangular first shell, without
requiring those directions to be commensurate with the square box. The mean
over all 24 annular values was the non-maximizing sensitivity. Both scores are
permutation invariant and exactly invariant to a common translation on the
allowed periodic reciprocal grid. Neither is exactly invariant to an arbitrary
rotation on the finite square reciprocal grid; selecting over the full annulus
and retaining its mean provide complementary directional sensitivities. They
contain no bond angles, ordered-mask membership, label values, or fitted
parameters. The primary analysis used both labels at all 45 sampled times and
four fixed leads. The timestep comparison used both timestep settings at the
24 post-quench times and four fixed leads. Dense
intervals used one two-score-by-time simultaneous family. Fixed-lead results
used an orientation-plus-translation $4\times4$ family containing raw
$\Psi_6$, the state score, $\Cord$, and $\sT$. Broader primary families added stored
energy and the annular mean; the timestep-comparison family contained the three
orientational scores and both translational scores. All uncertainty used 5000
complete-parent draws and equal duration weighting.

\subsection{Observable association, duration, and fixed-lead analyses}

Associations among current observables and with terminal structural quantities
used a duration-adjusted Spearman coefficient. Ranks were calculated separately
within each $\tau_q$, centered within that duration, and pooled only after
centering. Parent clusters were resampled for 5000 percentile draws. This
quantity is a finite-window association; it is not an equilibrium
autocorrelation function or relaxation-time estimate. Terminal targets were
the terminal-window mean of raw $\Psi_6$ for raw $\Psi_6$ and $\Cord$, the
corresponding structural state score, and the matched terminal
structural quantity for the other observables. Selected correlations and
quench-duration-specific fixed-lead results are reported below and in
Table~\ref{tab:s12couplingduration}.

At $t_{\rm post}=0.55$, the separately defined translational score $\sT$ had
duration-adjusted Spearman correlations of 0.872 with raw $\Psi_6$, 0.838 with
the state score, and 0.855 with $\Cord$. Their parent-cluster intervals were
$[0.852,0.891]$, $[0.812,0.862]$, and $[0.831,0.877]$, respectively
(Table~\ref{tab:s12couplingduration}). These associations show that the
translational comparison probes a strongly coupled aspect of the developing
collective order rather than a statistically independent signal.

Because the primary last-15\% label window has a duration-dependent absolute
width, its mean frame counts and starts were tabulated separately for each
$\tau_q$. A second label used a fixed final window of 0.50 time units. It
changed ordered-label counts from $(32,33,30,27)$ to $(28,31,32,30)$ over
$\tau_q=(0.15,0.60,1.50,3.00)$, with respective agreements 91.7\%, 95.0\%,
98.3\%, and 80.8\%. No duration was normalized by an independently measured
equilibrium structural-relaxation time.

The fixed-lead analysis was also repeated relative to the fixed-window start,
using the fixed-window outcome. At requested lead 1.0, raw $\Psi_6$, the state score,
and $\Cord$ had equal-duration AUROCs 0.551, 0.548, and 0.556. Their lower
limits under the energy-expanded four-observable-by-four-lead family were 0.461, 0.458,
and 0.466. At lead 0.50 the corresponding estimates were 0.597, 0.607, and
0.596, with joint lower limits 0.507, 0.517, and 0.507. Thus the fixed-window-label
analysis shows a rising fixed-lead profile, but its lead-1.0 lower limits include 0.5. At
lead 0.75, the three displayed lower limits round to 0.500, but the unrounded
values lie on both sides of chance; that row is not a joint three-observable
pass.

For the fixed-lead analysis, the target time for each trajectory was its primary
terminal-window start minus $\ell$, with
$\ell\in\{0.25,0.50,0.75,1.00\}$. The last stored frame not later than that
target was selected. Every selected frame strictly preceded the terminal
window; the minimum realized separation was 0.256. Within-duration AUROCs
and parent-cluster percentile intervals were then recomputed at each lead.
Every bootstrap draw used one common resample of the 40 parent clusters for
all leads and observables. In addition to pointwise intervals, an
observable-wise band controlled the four leads of one observable, and the
  primary joint band controlled all $3\times4$ estimates for raw $\Psi_6$, the
state score, and $\Cord$. As a broader sensitivity family, we also controlled
all $6\times4$ estimates after adding $\Cnb$, $\ford$, and $\Cmax$. Each band
used the 95th percentile of the bootstrap maximum absolute deviation from the
corresponding original-sample estimate. These families were specified before
their corrected intervals were inspected.

For the non-orientational comparison, the total potential
energy $U$ stored at every trajectory frame was converted to
$s_E=-U/(N\varepsilon)$. The sign was fixed so that a more negative potential
energy produces a larger ranking score. This scalar contains no bond angle,
ordered-mask, or fitted-model input. It was evaluated at the same 45 dense
times under both terminal labels. Pointwise parent-cluster intervals and
simultaneous lower bands were calculated over the full 45-time and
24-time post-quench families. We also evaluated the four fixed leads relative
to each label-window start. Those intervals used one joint family over raw
$\Psi_6$, the state score, $\Cord$, and $s_E$ at all four leads. This comparison
used 5000 parent-cluster draws and no trajectory generation or model fitting.

For the primary label at lead 1.0, $s_E$ had AUROC 0.489 with joint
$4\times4$ interval $[0.403,0.574]$. Raw $\Psi_6$, the state score, and $\Cord$ had
lower limits 0.513, 0.507, and 0.519 in the expanded family. At leads 0.50 and
0.25, however, $s_E$ had AUROCs 0.625 and 0.601, with joint intervals
$[0.539,0.710]$ and $[0.516,0.686]$; both lower limits exceeded 0.5 after the same simultaneous correction.
Its lead-0.75 interval, $[0.476,0.647]$, crossed 0.5. The primary-label energy
curve rose to 0.591 at $t_{\rm post}=1.20$, where the full 45-time lower
band was 0.494. Its lower band therefore did not remain above 0.5 in the full
45-time analysis; in the 24-time post-quench analysis this occurred only at the
final 1.20 grid point. Under the fixed-window label, neither lower band remained
above 0.5 through 1.20. Figure~\ref{fig:s17} and
Table~\ref{tab:s7energy} report all estimates.

The primary translational-order score changed the specificity interpretation. In
the primary analysis, $\sT$ had AUROC 0.593 at lead 1.0; its
orientation-plus-translation $4\times4$ interval was $[0.511,0.675]$, and its
broader all-six interval was $[0.502,0.684]$. The fixed-window-label value was
0.560 $[0.475,0.645]$ in the $4\times4$ family, and the fine-step value was
0.539 $[0.454,0.625]$; both intervals included 0.5 at lead 1.0. At lead 0.50, the respective
values were 0.673 $[0.591,0.754]$, 0.616 $[0.531,0.700]$, and 0.627
$[0.542,0.713]$, all above chance after the same joint correction. They also
had lower limits above 0.5 in each analysis's broader family. The annular mean
rose in all three analyses, but its broadest fixed-window interval at
lead 0.50 had an unrounded lower limit of 0.4990 and therefore did not pass.

For $\sT$, the earliest persistently discriminative sampled times under the
dense two-score joint bands were 0.55 for the primary analysis and 0.85 for both
the fixed-window-label and fine-step analyses. The corresponding annular-mean
values were 0.55, 1.10, and 0.85. No paired fine-step-minus-primary dense or
fixed-lead translational interval excluded zero after simultaneous correction.
Figure~\ref{fig:s18} and Table~\ref{tab:s8translation} report the complete
summary. The common top-three result at lead 0.50 supports structural-order
discrimination, whereas the variation at lead 1.0 and between translational
definitions precludes an orientation-specific or uniquely timed claim.

The quench-duration-specific comparison kept the four durations as separate cells. It
reported pointwise parent-cluster intervals, one simultaneous family over the
$4\times4$ duration-by-lead cells of each observable, and a joint family over
all $3\times4\times4$ primary orientational cells. At requested lead 1.0, the selected
post-quench times were 0.642, 0.576, 0.460, and 0.224 for
$\tau_q\in\{0.15,0.60,1.50,3.00\}$. All three pointwise lower limits exceeded
0.5 only for $\tau_q=0.15$; no duration had this property under the
observable-wise 16-cell correction or the stricter 48-cell correction
(Fig.~\ref{fig:s15}). Thus fixed lead holds distance to the label window
approximately constant but does not hold post-quench age constant.

We then added $\sT$ to form the four-measure duration-resolved family reported
in Table~\ref{tab:s12couplingduration}. At lead 0.50, both the pointwise and
observable-wise 16-cell lower limits exceeded 0.5 for all four measures at
$\tau_q=0.15$ and 0.60.
For $\tau_q=1.50$ and 3.00, one pointwise lower limit exceeded 0.5 at each
duration, but none of the four observable-wise lower limits did so. Under the
joint 64-cell family, three of four lower limits exceeded 0.5 at $\tau_q=0.15$,
all four did so at 0.60, and none did so at 1.50 or 3.00. The equal-duration
headline therefore summarizes a signal that is resolved most consistently for
the two shorter quench durations in the present sample.

\subsection{Definition and fixed-cohort sensitivity analyses}

The primary neighbor-phase coherence normalized each nonzero local complex
order parameter before taking the real phase product over neighbor edges. Its
endpoint-amplitude cutoff was varied from $10^{-12}$ to 0.1 or 0.3. An
amplitude-weighted alternative used edge weights
$|\psi_{6,i}||\psi_{6,j}|$. Point curves and 24-time post-quench simultaneous
bands were recomputed for all four definitions from the same trajectories.

The cohort analysis fixed membership once at p050 by excluding trajectories that
already satisfied both instantaneous structural thresholds. The resulting 393
trajectories, including 95 ordered outcomes, were followed unchanged at p050
and p100. State and baseline GNN scores used their original folds and
predictions. Both parent-cluster and hierarchical seed--parent bootstrap intervals were applied,
so this analysis does not condition on a different subset at p100. At p050,
the fixed and phase-specific definitions select exactly the same trajectories
and state statistic. Table~\ref{tab:s3results} therefore reuses the
phase-specific state interval instead of presenting the immaterial
$10^{-3}$-level difference from an independent bootstrap realization. In a separate
q25 comparison, the apparent below-chance state AUROC was not distinguished
from chance because its upper parent-cluster interval exceeded 0.5.

Finally, each terminal-label variant was checked at the actual
seed--fold--duration cell level. A cell was valid only if both labels were
present. Because strict variants retained as few as 34/60 or 41/60 cells, their
within-duration AUROCs are evaluated on smaller domains and are not directly
precision-matched to the 60/60-cell variants. These sensitivity analyses are summarized in
Fig.~\ref{fig:s12} and Table~\ref{tab:s5labels}.

\subsection{First-crossing time and terminal structure}

The auxiliary $t^*$ retained its training-time definition: the first stored
time at which a trailing window of 7.5\% of the trajectory, with a minimum of
two frames, simultaneously met mean $\Psi_6\ge0.55$ and mean
$\ford\ge0.75$. It was analyzed as a first crossing without an absorbing-state
assumption. A finite $t^*$ followed by a terminal non-ordered label was counted
as a transient crossing. No right-censored survival model was fitted because
the physical event was not shown to be absorbing.

For every transient crossing, the three terminal criteria were evaluated
separately: terminal-window mean $\Psi_6<0.55$, terminal-window mean
$\ford<0.75$, and terminal-window standard deviation of $\Psi_6>0.08$.
Marginal failure fractions count a trajectory in every failed criterion, whereas
the exclusive breakdown records the exact combination. Parent-cluster draws
resampled parent clusters and recomputed each fraction among the transient
crossings retained in that draw. The marginal categories therefore need not sum
to one.

Terminal pair structure was averaged over the same primary label window. For
particle pairs separated by $r$, $g(r)$ used ideal-gas shell normalization in
the periodic square cell, and the conditional orientational curve was
\begin{equation}
 G_6(r)=\left\langle
 \operatorname{Re}\!\left[\psi_{6,i}\psi_{6,j}^{*}\right]
 \middle|\ r_{ij}\ \text{in the bin around }r
 \right\rangle .
\end{equation}
Both curves used 48 bins from zero to $L/2$. Coordination fractions for
$z=5,6,7$, and values outside 5--7 were calculated with $r_{\rm nb}=1.5$.
Outcome-specific means and 95\% intervals used complete-parent resampling. We also
averaged each coordination fraction within parent and outcome. Forty parents
contributed a non-ordered outcome, 39 an ordered outcome, and 39 contributed
both. For the paired set, each bootstrap draw resampled parents and averaged the
within-parent ordered-minus-non-ordered difference (Fig.~\ref{fig:s16}). The
terminal fraction of six-coordinated particles was the mean, over the primary terminal window, of
$N^{-1}\sum_i\mathbf{1}(n_i=6)$. It is not a direct term
in the binary terminal label, although it is physically related to the same
ordering process. At each fixed lead, its association with current raw
$\Psi_6$, state score, and $\Cord$ was quantified by the duration-adjusted
Spearman coefficient defined above. One common parent resample was used over
the $3\times4$ association family, and simultaneous intervals were constructed
from its maximum absolute bootstrap deviation. We treat this endpoint as an
exploratory continuous structural check, not as an independent mechanism or a
redefined phase label. The
terminal snapshots in Fig.~\ref{fig:s13} show the three replicas from one mixed
replica set at their final stored frame.

\subsection{Numerical and implementation checks}

\begin{figure}[p]
\centering
\includegraphics[width=\textwidth]{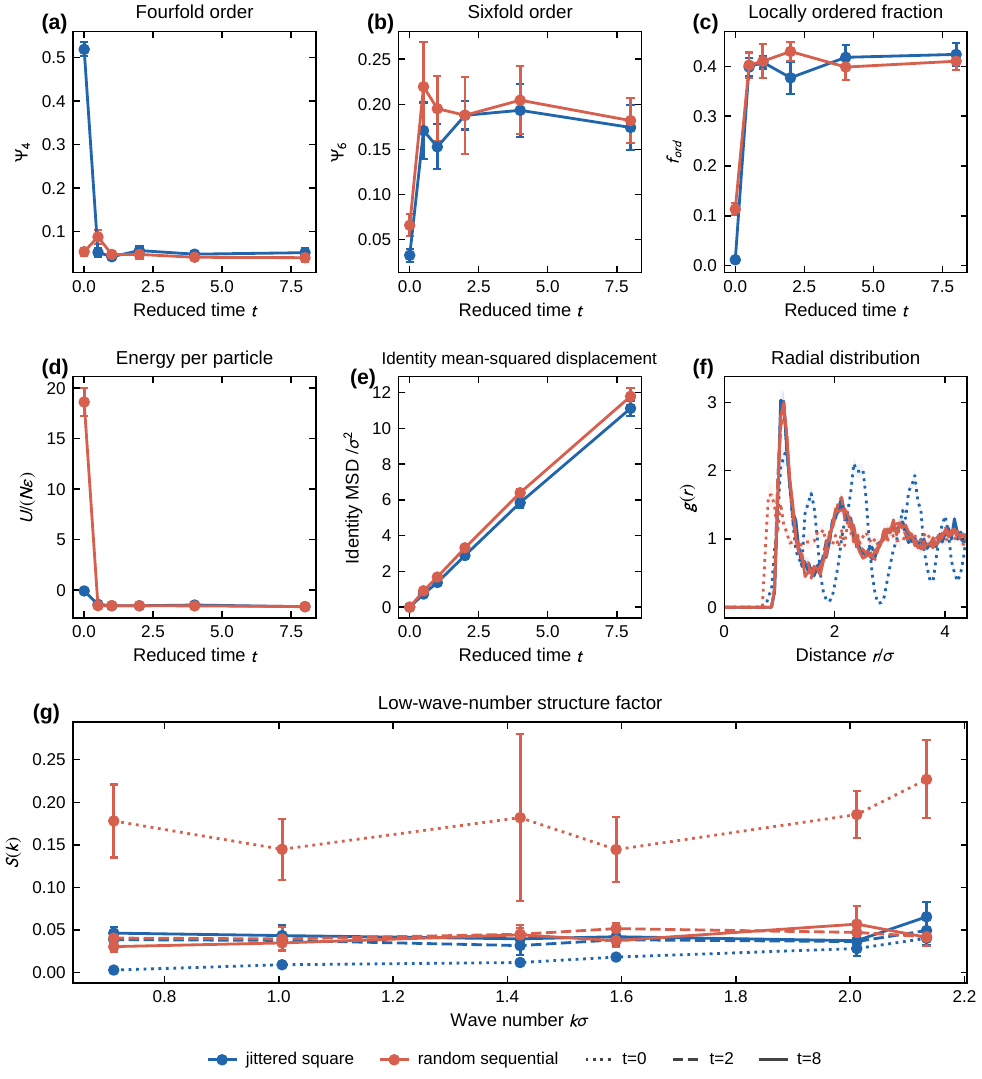}
\caption{\textbf{High-temperature burn-in and initializer comparison.} (a) Global
$\Psi_4$, (b) global $\Psi_6$, (c) locally ordered fraction, (d) energy per
particle, and (e) identity mean-squared displacement through time 8. Points and
bars show family means and standard errors across eight seeds. (f) Radial
distribution function and (g) low-wave-number structure factor at times 0, 2,
and 8 for jittered-square and random-sequential initializations. Shading or
error bars show standard errors across seeds. Random-sequential states begin
at higher but finite energy and relax rapidly; no step crosses the predefined
$E/N=100$ numerical-instability threshold. Energy, displacement, distance, and
wave number are plotted as $U/(N\varepsilon)$, ${\rm MSD}/\sigma^2$,
$r/\sigma$, and $k\sigma$, respectively}
\label{fig:s7}
\end{figure}

The high-temperature initializer comparison followed eight weakly perturbed
square states and eight random-sequential states through time 8 at $T=1.8$.
Both families relaxed without runaway energies, although the random-sequential
states began at substantially higher energy (Fig.~\ref{fig:s7}).

The fine-step ensemble repeated the 40-parent, four-duration, three-replica
design with $\Delta t=2\times10^{-5}$ instead of $5\times10^{-5}$. Both
ensembles stored frames every 0.016 time units and evaluated the same force and
energy functions. Direct evaluation on identical stored coordinates agreed to
floating-point precision. Because the timestep changes the Brownian-noise grid,
the trajectory comparison is distributional rather than pathwise; two step
sizes also do not constitute a continuum extrapolation.

\begin{figure}[p]
\centering
\includegraphics[width=\textwidth]{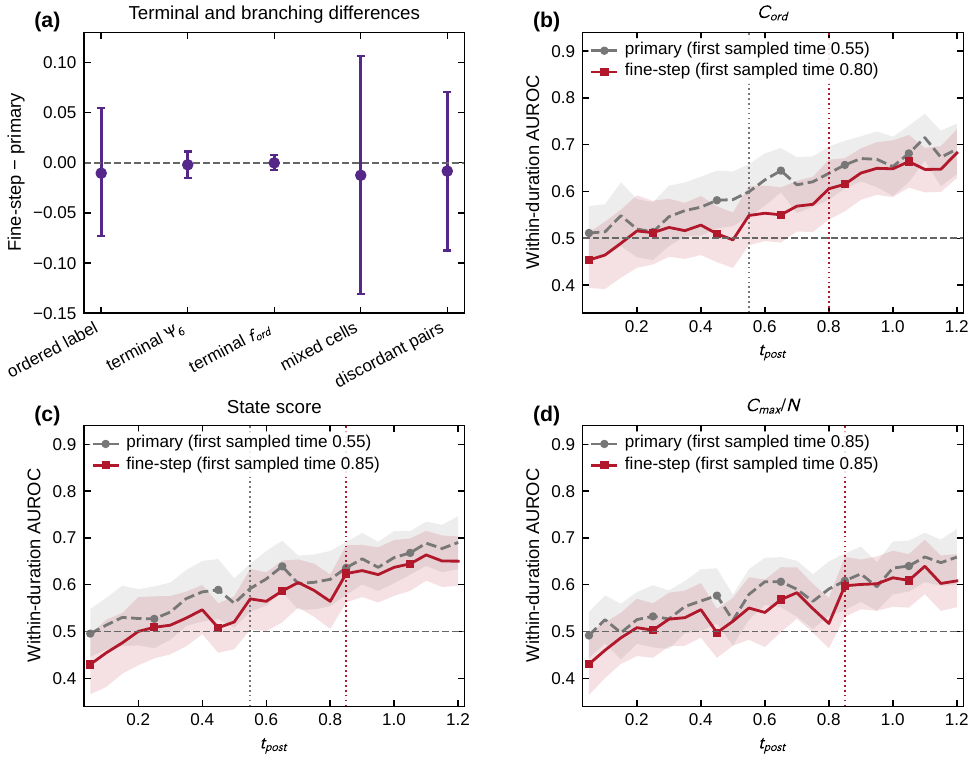}
\caption{\textbf{Parent-matched timestep comparison.} (a) Parent-cluster
intervals for paired fine-step-minus-primary changes in terminal and branching
quantities. The ordered-label rate and the two terminal structural means show
small paired differences; the branching contrasts are shown alongside them.
(b--d) Dense post-quench AUROC with pointwise intervals for
$\Cord$, state score, and $\Cmax$. Vertical lines mark the earliest persistently
discriminative sampled times under the observable-wise bands. The primary and
fine-step ensembles use different timesteps and
Brownian-noise grids. Their different sampled times do not establish temporal
convergence}
\label{fig:s8}
\end{figure}

Post-quench structural discrimination was recalculated over 24 times from
$t_{\rm post}=0.05$ to 1.20 and at four fixed leads. Each ensemble used its own
terminal outcomes and terminal-window starts. In the fine-step ensemble, raw
$\Psi_6$, the state score, and $\Cord$ had AUROCs 0.546, 0.542, and 0.544 at
lead 1.0, with simultaneous intervals crossing 0.5. At lead 0.75 the estimates
were 0.614, 0.615, and 0.605, with lower limits 0.533, 0.533, and 0.523. Paired
fine-step-minus-primary intervals included zero throughout the corrected
families. The comparison therefore reproduces the development of structural
outcome association without defining a continuum onset (Figs.~\ref{fig:s8}
and~\ref{fig:s12}). Exact software requirements and random-number settings are
provided with the accompanying reproducibility package.

\section{Supplementary tables}

\begin{table}[htbp]
\centering
\caption{Primary and comparison ensembles. Parents are indivisible grouping blocks; the primary ensemble also supplies the primary-timestep reference}
\label{tab:s1design}
\small
\renewcommand{\arraystretch}{1.10}
\setlength{\tabcolsep}{2.5pt}
\begin{tabular}{@{}>{\raggedright\arraybackslash}p{0.20\linewidth}c>{\raggedright\arraybackslash}p{0.15\linewidth}>{\raggedright\arraybackslash}p{0.11\linewidth}>{\raggedright\arraybackslash}p{0.13\linewidth}>{\centering\arraybackslash}p{0.08\linewidth}>{\centering\arraybackslash}p{0.07\linewidth}>{\centering\arraybackslash}p{0.11\linewidth}@{}}
\toprule
Role & $N$ & Temperature(s) & $\tau_q$ & Parent blocks & \shortstack{Replicas\\per set} & Runs & $\Delta t$ \\
\midrule
Primary ensemble & 64 & $T_{\rm low}=0.50$ & four durations & 40 & 3 & 480 & $5\times10^{-5}$ \\
Fine-step ensemble & 64 & $T_{\rm low}=0.50$ & four durations & 40 & 3 & 480 & $2\times10^{-5}$ \\
Initializer comparison, each family & 64 & $T=1.8$, no quench & --- & 8 per initializer & --- & 8 per initializer & $5\times10^{-5}$ \\
Independent larger-system ensemble & 128 & $T_{\rm low}=0.35,0.50,0.65$ & four durations & 20 shared across 12 conditions & 1 & 240 & $2\times10^{-5}$ \\
\bottomrule
\end{tabular}
\end{table}

\clearpage
\begin{landscape}
\begin{table}[H]
\centering
\caption{Summary of primary and timestep-sensitivity results. Intervals are
95\% parent-cluster bootstrap intervals unless a simultaneous family, a second
uncertainty level, or a paired contrast is explicitly named}
\label{tab:s3results}
\scriptsize
\setlength{\tabcolsep}{3pt}
\renewcommand{\arraystretch}{0.87}
\begin{tabularx}{\linewidth}{@{}>{\raggedright\arraybackslash}p{0.18\linewidth}>{\raggedright\arraybackslash}p{0.10\linewidth}>{\raggedright\arraybackslash}p{0.13\linewidth}>{\raggedright\arraybackslash}p{0.18\linewidth}>{\raggedright\arraybackslash}p{0.21\linewidth}>{\raggedright\arraybackslash}X@{}}
\toprule
Analysis & Sampled time & Estimate & Parent interval or sampled time & Second-level or paired result & Interpretation \\
\midrule
Mixed replica sets & terminal & 92/160 & --- & Bernoulli expectation: 90.8 & matches duration-wise Bernoulli expectation \\
Corrected between-parent variance & all $\tau_q$ & $-0.0009$ & $[-0.0180,0.0139]$ & $f_{\rm parent}$ 97.5th percentile: 7.5\% & compatible with zero at $m=3$ \\
Raw $\Psi_6$/state/$\Cord$: first times & 45 times & 0.55/0.60/0.55 & joint: 0.60/0.80/0.60 & 24-time: 0.55/0.55/0.55 & depends on multiplicity family \\
$\Cnb$/$\ford$/$\Cmax$: first times & 45 times & 0.80/0.85/1.00 & joint: 0.80/1.00/1.00 & 24-time: 0.55/0.80/0.80 & $\Cmax$ observable-wise entries use $100\,000$ draws; others use 5000 \\
Raw $\Psi_6$/state/$\Cord$ & fixed lead 1.0 & 0.599/0.593/0.605 & \shortstack[l]{joint $3\times4$:\\$[0.519,0.678]$\\$[0.513,0.672]$\\$[0.525,0.684]$} & \shortstack[l]{joint $6\times4$ lower limits:\\0.508/0.502/0.514} & aggregate lower limits exceed 0.5 after equal duration weighting; not established within every duration \\
Raw $\Psi_6$/state/$\Cord$ versus terminal six-coordinated fraction & fixed lead 1.0 & 0.168/0.156/0.202 & \shortstack[l]{joint $3\times4$:\\$[0.060,0.276]$\\$[0.048,0.265]$\\$[0.094,0.310]$} & duration-adjusted Spearman $\rho$ & exploratory continuous endpoint \\
Fine-step raw $\Psi_6$/state/$\Cord$ & fixed lead 1.0 & 0.546/0.542/0.544 & \shortstack[l]{joint $3\times4$:\\$[0.464,0.627]$\\$[0.460,0.623]$\\$[0.463,0.626]$} & \shortstack[l]{paired fine-step--primary:\\all corrected intervals include zero} & intervals include 0.5 at lead 1.0 \\
Fine-step raw $\Psi_6$/state/$\Cord$ & fixed lead 0.75 & 0.614/0.615/0.605 & \shortstack[l]{joint $3\times4$:\\$[0.533,0.696]$\\$[0.533,0.696]$\\$[0.523,0.686]$} & paired fine-step--primary intervals include zero & lower limits exceed 0.5 at lead 0.75 \\
State score & p050 full cohort & 0.564 & $[0.513,0.605]$ & $[0.504,0.615]$ & both lower limits exceed 0.5; repeat is the fold partition \\
Baseline GNN & p100 & 0.624 & $[0.560,0.684]$ & $[0.525,0.710]$ & both lower limits exceed 0.5 \\
Duration-conditioned GNN & p100 & 0.653 & $[0.592,0.707]$ & $[0.579,0.719]$ & hierarchical seed--parent contrast includes zero \\
Targeted GNN & p050 & 0.588 & $[0.529,0.635]$ & $[0.507,0.653]$ & both lower limits exceed 0.5 \\
State/GNN, cohort defined at p050 & p050 & \shortstack[l]{0.551\\0.500} & \shortstack[l]{$[0.486,0.596]$\\$[0.450,0.535]$} & \shortstack[l]{$[0.469,0.611]$\\$[0.408,0.578]$} & state interval reused from the identical phase-specific cohort; intervals include 0.5 \\
State/GNN, cohort defined at p050 & p100 & \shortstack[l]{0.657\\0.631} & \shortstack[l]{$[0.595,0.722]$\\$[0.558,0.699]$} & \shortstack[l]{$[0.581,0.734]$\\$[0.509,0.733]$} & all lower limits exceed 0.5 \\
Finite $t^*$ among non-ordered outcomes & terminal & 219/358 & --- & $207/219$ do not satisfy terminal $\ford$ criterion & crossing is nonabsorbing; the $\ford$ criterion is the most frequently unsatisfied marginal criterion \\
\bottomrule
\end{tabularx}
\end{table}
\end{landscape}

\clearpage
\begin{table}[H]
\centering
\caption{Terminal-label sensitivity with all predictor scores held fixed.
Only the terminal outcomes change; the state score retains denominators 0.55
and 0.75, and stored GNN predictions are not retrained. ``Valid cells'' is the
number of seed--fold--duration cells containing both classes at each phase (60
possible). Bold
AUROCs meet the full above-chance criterion: the point estimate and the lower limits
of both parent-cluster and hierarchical seed--parent bootstrap intervals exceed 0.5. The state score
passes at p100 for 10/11 variants but at p050 for 5/11; the baseline GNN
passes at p100 for 8/11. Here $w$ is the fractional terminal-window length}
\label{tab:s5labels}
\small
\resizebox{\linewidth}{!}{%
\begin{tabular}{@{}lccccrrccc@{}}
\toprule
Variant & $\Psi_6$ cutoff & $\ford$ cutoff & stability cutoff & $w$ & $n_+$ & valid cells & state p050 & state p100 & GNN p100 \\
\midrule
baseline & 0.55 & 0.75 & 0.08 & 0.15 & 122 & 60/60 & \textbf{0.564} & \textbf{0.656} & \textbf{0.624} \\
relaxed joint & 0.50 & 0.70 & 0.10 & 0.15 & 265 & 60/60 & 0.564 & \textbf{0.657} & \textbf{0.612} \\
strict joint & 0.60 & 0.80 & 0.06 & 0.15 & 15 & 34/60 & 0.391 & 0.597 & 0.497 \\
$\Psi_6$ loose & 0.50 & 0.75 & 0.08 & 0.15 & 125 & 60/60 & \textbf{0.561} & \textbf{0.654} & \textbf{0.626} \\
$\Psi_6$ strict & 0.60 & 0.75 & 0.08 & 0.15 & 116 & 60/60 & \textbf{0.564} & \textbf{0.657} & \textbf{0.633} \\
$\ford$ loose & 0.55 & 0.70 & 0.08 & 0.15 & 217 & 60/60 & \textbf{0.574} & \textbf{0.655} & \textbf{0.605} \\
$\ford$ strict & 0.55 & 0.80 & 0.08 & 0.15 & 22 & 41/60 & 0.446 & \textbf{0.621} & 0.530 \\
stability loose & 0.55 & 0.75 & 0.10 & 0.15 & 129 & 60/60 & \textbf{0.561} & \textbf{0.660} & \textbf{0.624} \\
stability strict & 0.55 & 0.75 & 0.06 & 0.15 & 91 & 58/60 & 0.545 & \textbf{0.640} & \textbf{0.617} \\
window short & 0.55 & 0.75 & 0.08 & 0.10 & 120 & 60/60 & 0.536 & \textbf{0.598} & 0.576 \\
window long & 0.55 & 0.75 & 0.08 & 0.20 & 113 & 60/60 & 0.561 & \textbf{0.710} & \textbf{0.640} \\
\bottomrule
\end{tabular}%
}
\end{table}

\clearpage
\begin{table}[H]
\centering
\caption{Potential-energy and fixed-window fixed-lead comparison. Each entry is
the equal-duration AUROC followed by its joint $4\times4$ parent-cluster
interval in brackets. The family contains raw $\Psi_6$, the state score, $\Cord$, and
$-U/(N\varepsilon)$ at four leads. All intervals use 5000 parent-cluster
draws. Lower limits are compared with 0.5 before rounding; a displayed lower limit
of 0.500 need not exceed 0.5 on the unrounded scale}
\label{tab:s7energy}
\scriptsize
\setlength{\tabcolsep}{3.4pt}
\resizebox{\linewidth}{!}{%
\begin{tabular}{@{}llcccc@{}}
\toprule
Terminal label & Lead & Raw $\Psi_6$ & State & $\Cord$ & $-U/(N\varepsilon)$ \\
\midrule
Primary fractional & 1.00 & 0.599 [0.513, 0.684] & 0.593 [0.507, 0.678] & 0.605 [0.519, 0.690] & 0.489 [0.403, 0.574] \\
Primary fractional & 0.75 & 0.647 [0.562, 0.732] & 0.628 [0.542, 0.713] & 0.655 [0.570, 0.740] & 0.562 [0.476, 0.647] \\
Primary fractional & 0.50 & 0.660 [0.575, 0.746] & 0.647 [0.562, 0.733] & 0.646 [0.560, 0.731] & 0.625 [0.539, 0.710] \\
Primary fractional & 0.25 & 0.685 [0.599, 0.770] & 0.655 [0.570, 0.740] & 0.701 [0.616, 0.787] & 0.601 [0.516, 0.686] \\
\midrule
Fixed final 0.50 & 1.00 & 0.551 [0.461, 0.641] & 0.548 [0.458, 0.638] & 0.556 [0.466, 0.646] & 0.518 [0.428, 0.607] \\
Fixed final 0.50 & 0.75 & 0.590 [0.500, 0.680] & 0.590 [0.500, 0.679] & 0.590 [0.500, 0.680] & 0.557 [0.467, 0.646] \\
Fixed final 0.50 & 0.50 & 0.597 [0.507, 0.687] & 0.607 [0.517, 0.697] & 0.596 [0.507, 0.686] & 0.557 [0.468, 0.647] \\
Fixed final 0.50 & 0.25 & 0.692 [0.602, 0.782] & 0.683 [0.593, 0.773] & 0.690 [0.600, 0.780] & 0.585 [0.495, 0.675] \\
\bottomrule
\end{tabular}%
}
\end{table}

\clearpage
\begin{table}[H]
\centering
\caption{First-reciprocal-shell translational-order comparison. Fixed-lead entries are
equal-duration AUROCs followed by the broadest available joint interval in
brackets. For the two primary-timestep rows, the family contains raw $\Psi_6$,
the state score, $\Cord$, $-U/(N\varepsilon)$, $\sT$, and the annular mean at four leads
($6\times4$). For the fine-step rows it contains the three orientational and
two translational scores at four leads ($5\times4$). The final column is the
earliest post-quench sampled time whose lower band remains above 0.5 through 1.20 in
the joint two-translational-score dense family. A value is a sampled time,
not an exact onset. All intervals use 5000 parent-cluster draws}
\label{tab:s8translation}
\small
\setlength{\tabcolsep}{4.0pt}
\resizebox{\linewidth}{!}{%
\begin{tabular}{@{}llccc@{}}
\toprule
Analysis & Translational-order score & Lead 1.00 & Lead 0.50 & Dense joint sampled time \\
\midrule
Primary analysis & Top-three $\sT$ & 0.593 [0.502, 0.684] & 0.673 [0.582, 0.763] & 0.55 \\
Primary analysis & Annular mean & 0.583 [0.492, 0.674] & 0.636 [0.545, 0.727] & 0.55 \\
Fixed-window-label analysis & Top-three $\sT$ & 0.560 [0.466, 0.654] & 0.616 [0.521, 0.710] & 0.85 \\
Fixed-window-label analysis & Annular mean & 0.568 [0.474, 0.663] & 0.593 [0.499, 0.688] & 1.10 \\
Fine-step analysis & Top-three $\sT$ & 0.539 [0.450, 0.629] & 0.627 [0.538, 0.717] & 0.85 \\
Fine-step analysis & Annular mean & 0.577 [0.488, 0.667] & 0.610 [0.521, 0.700] & 0.85 \\
\bottomrule
\end{tabular}%
}
\end{table}

\clearpage
\begin{landscape}
\begin{table}[H]
\centering
\caption{Fixed-lead results under the joint
$4\times4$ orientation-plus-translation family. Each entry is the
equal-duration AUROC followed by its simultaneous 95\% parent-cluster
interval. The family contains raw $\Psi_6$, the structural state score,
$\Cord$, and $\sT$ at all four leads. All intervals use 5000 parent-cluster
draws}
\label{tab:s11fixedlead4x4}
\small
\setlength{\tabcolsep}{4.4pt}
\renewcommand{\arraystretch}{1.05}
\begin{tabular}{@{}llcccc@{}}
\toprule
Analysis & Lead & Raw $\Psi_6$ & State score & $\Cord$ & $\sT$ \\
\midrule
Primary analysis & 1.00 & 0.599 [0.517, 0.680] & 0.593 [0.511, 0.674] & 0.605 [0.523, 0.686] & 0.593 [0.511, 0.675] \\
 & 0.75 & 0.647 [0.565, 0.729] & 0.628 [0.546, 0.710] & 0.655 [0.573, 0.737] & 0.637 [0.555, 0.719] \\
 & 0.50 & 0.660 [0.578, 0.742] & 0.647 [0.566, 0.729] & 0.646 [0.564, 0.727] & 0.673 [0.591, 0.754] \\
 & 0.25 & 0.685 [0.603, 0.766] & 0.655 [0.573, 0.737] & 0.701 [0.620, 0.783] & 0.703 [0.621, 0.785] \\
\midrule
Fixed-window-label analysis & 1.00 & 0.551 [0.466, 0.636] & 0.548 [0.463, 0.633] & 0.556 [0.471, 0.641] & 0.560 [0.475, 0.645] \\
 & 0.75 & 0.590 [0.505, 0.675] & 0.590 [0.505, 0.674] & 0.590 [0.505, 0.675] & 0.612 [0.527, 0.697] \\
 & 0.50 & 0.597 [0.512, 0.682] & 0.607 [0.522, 0.692] & 0.596 [0.511, 0.681] & 0.616 [0.531, 0.700] \\
 & 0.25 & 0.692 [0.607, 0.777] & 0.683 [0.598, 0.768] & 0.690 [0.605, 0.775] & 0.691 [0.606, 0.776] \\
\midrule
Fine-step analysis & 1.00 & 0.546 [0.460, 0.631] & 0.542 [0.456, 0.627] & 0.544 [0.459, 0.630] & 0.539 [0.454, 0.625] \\
 & 0.75 & 0.614 [0.529, 0.700] & 0.615 [0.529, 0.700] & 0.605 [0.519, 0.690] & 0.619 [0.534, 0.705] \\
 & 0.50 & 0.618 [0.532, 0.703] & 0.598 [0.512, 0.683] & 0.626 [0.541, 0.712] & 0.627 [0.542, 0.713] \\
 & 0.25 & 0.732 [0.646, 0.817] & 0.717 [0.632, 0.803] & 0.704 [0.619, 0.790] & 0.731 [0.645, 0.816] \\
\bottomrule
\end{tabular}
\end{table}
\end{landscape}

\clearpage
\begin{table}[H]
\centering
\caption{Translational--orientational coupling and duration-resolved results at
lead 0.50. Part (a) gives duration-adjusted Spearman correlations at
$t_{\rm post}=0.55$ with parent-cluster percentile intervals. In part (b),
``16-cell lower'' is the simultaneous lower limit over four durations and four
leads for one observable; ``64-cell lower'' is the simultaneous lower limit
over all four observables, durations, and leads. The same 5000 complete-parent
resamples were shared within each family}
\label{tab:s12couplingduration}
\small
\setlength{\tabcolsep}{4.5pt}
\renewcommand{\arraystretch}{1.00}
\textbf{(a) Translational--orientational correlations}\par\smallskip
\begin{tabular}{@{}lcc@{}}
\toprule
Observable paired with $\sT$ & Spearman $\rho$ & Parent-cluster 95\% interval \\
\midrule
Raw $\Psi_6$ & 0.872 & [0.852, 0.891] \\
State score & 0.838 & [0.812, 0.862] \\
$\Cord$ & 0.855 & [0.831, 0.877] \\
\bottomrule
\end{tabular}

\medskip
\textbf{(b) Quench-duration-specific AUROC at lead 0.50}\par\smallskip
\begin{tabular}{@{}llcccc@{}}
\toprule
$\tau_q$ & Observable & AUROC & Pointwise 95\% interval & 16-cell lower & 64-cell lower \\
\midrule
0.15 & Raw $\Psi_6$ & 0.714 & [0.585, 0.823] & 0.537 & 0.522 \\
 & State score & 0.693 & [0.564, 0.804] & 0.521 & 0.501 \\
 & $\Cord$ & 0.685 & [0.561, 0.788] & 0.514 & 0.493 \\
 & $\sT$ & 0.746 & [0.636, 0.842] & 0.573 & 0.554 \\
\midrule
0.60 & Raw $\Psi_6$ & 0.703 & [0.595, 0.805] & 0.525 & 0.510 \\
 & State score & 0.699 & [0.594, 0.802] & 0.526 & 0.506 \\
 & $\Cord$ & 0.705 & [0.603, 0.802] & 0.534 & 0.513 \\
 & $\sT$ & 0.726 & [0.629, 0.823] & 0.552 & 0.534 \\
\midrule
1.50 & Raw $\Psi_6$ & 0.619 & [0.489, 0.743] & 0.442 & 0.427 \\
 & State score & 0.618 & [0.504, 0.724] & 0.445 & 0.425 \\
 & $\Cord$ & 0.594 & [0.470, 0.716] & 0.423 & 0.402 \\
 & $\sT$ & 0.567 & [0.443, 0.690] & 0.393 & 0.375 \\
\midrule
3.00 & Raw $\Psi_6$ & 0.605 & [0.490, 0.715] & 0.428 & 0.413 \\
 & State score & 0.580 & [0.462, 0.702] & 0.407 & 0.387 \\
 & $\Cord$ & 0.598 & [0.484, 0.706] & 0.427 & 0.405 \\
 & $\sT$ & 0.651 & [0.548, 0.755] & 0.477 & 0.459 \\
\bottomrule
\end{tabular}
\end{table}

\clearpage
\section{Supplementary figures}

Unless stated otherwise, all time quantities in the supplementary figures are
reported in the reduced units defined in the main text.

\begin{figure}[p]
\centering
\includegraphics[width=\textwidth]{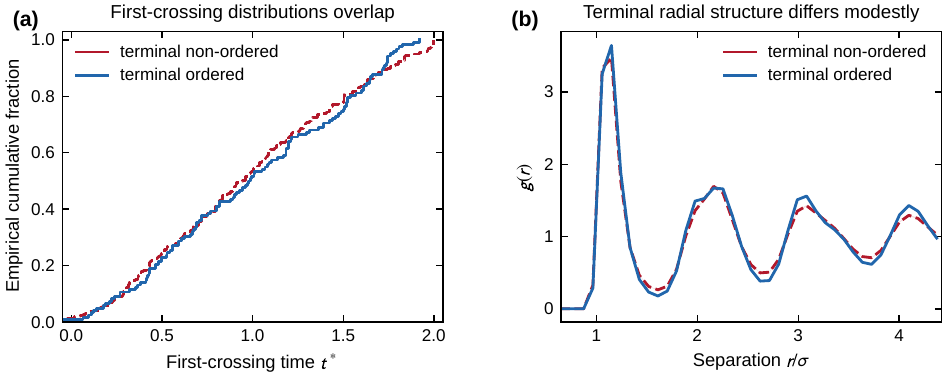}
\caption{\textbf{First crossings and terminal radial structure.}
(a) Empirical cumulative distributions of finite first-crossing times relative
to quench completion. The outcome distributions overlap, and 219 of 358
terminal non-ordered trajectories cross transiently. (b) Terminal-window
radial distribution $g(r)$ by operational outcome. Curves are means and 95\%
parent-cluster intervals; the differences are modest. Separation is reported
as $r/\sigma$}
\label{fig:s11}
\end{figure}

\begin{figure}[p]
\centering
\includegraphics[width=\textwidth]{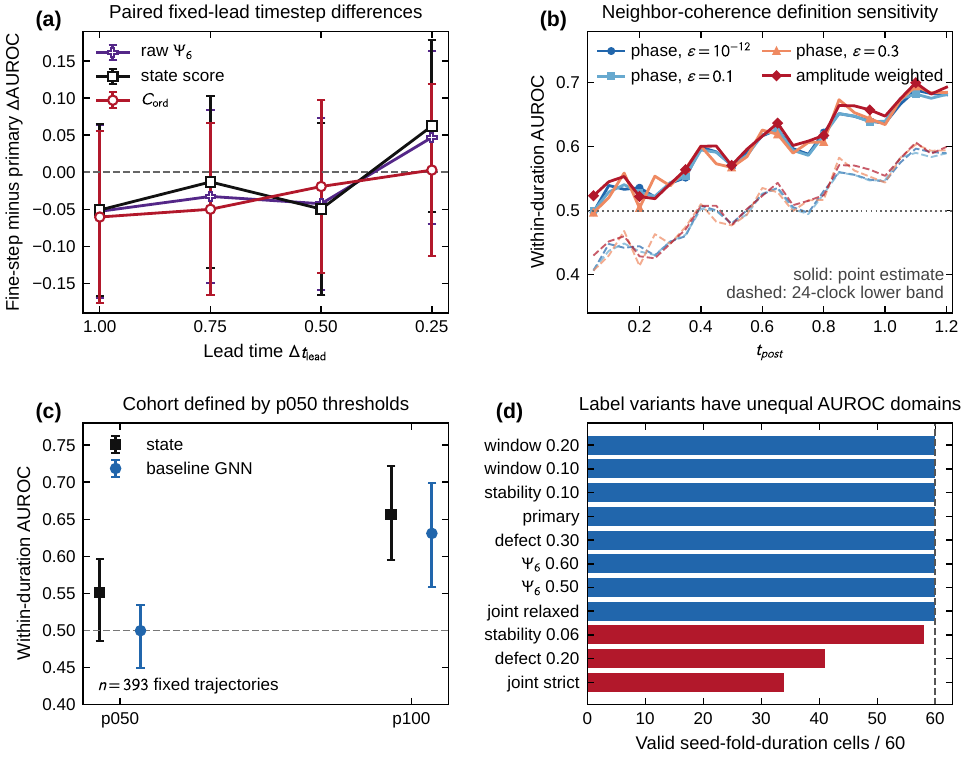}
\caption{\textbf{Timestep and structural-definition sensitivity analyses.}
(a) Paired fine-step-minus-primary fixed-lead AUROC differences for raw $\Psi_6$,
the state score, and $\Cord$. Error bars are joint $3\times4$
parent-cluster intervals; none excludes zero. (b) Neighbor-phase-coherence
curves under three endpoint-amplitude cutoffs and an amplitude-weighted
definition. Solid curves are point estimates, and dashed curves are 24-time
post-quench observable-wise lower bands. (c) State-score and baseline-GNN AUROC
for the cohort defined at p050 by excluding trajectories that already satisfy
both instantaneous thresholds and then followed unchanged to p100; error bars
show parent-cluster intervals, and the learned-model comparison also uses the
hierarchical seed--parent bootstrap interval reported in Table~\ref{tab:s3results}. (d) Number of
valid seed--fold--duration cells for each explicitly labeled terminal-window or
threshold variant. Strict variants
change the AUROC evaluation domain. The q25 state estimate is not distinguished
from chance because both relevant upper limits exceed 0.5}
\label{fig:s12}
\end{figure}

\begin{figure}[p]
\centering
\includegraphics[width=\textwidth]{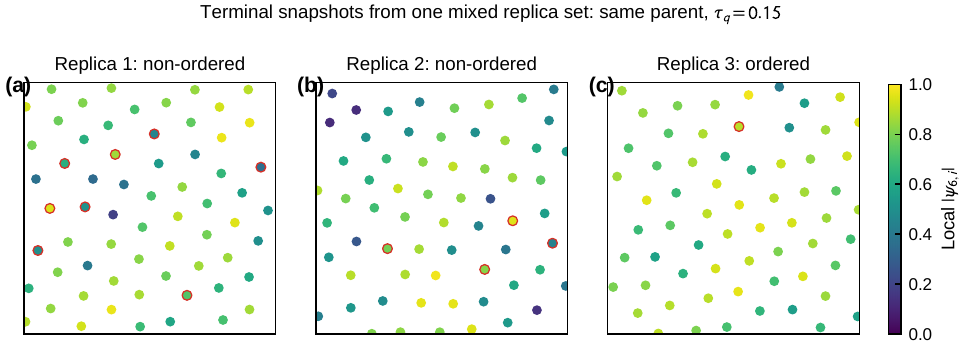}
\caption{\textbf{Terminal snapshots from one mixed replica set.} The three replicas share the same stored parent and
$\tau_q=0.15$ but use independent Brownian-noise realizations. Color gives
local $|\psi_{6,i}|$; red particle outlines mark coordination outside 5--7.
Each panel uses its last stored terminal frame and spans $L=8.8345\sigma$}
\label{fig:s13}
\end{figure}

\begin{figure}[p]
\centering
\includegraphics[width=\textwidth]{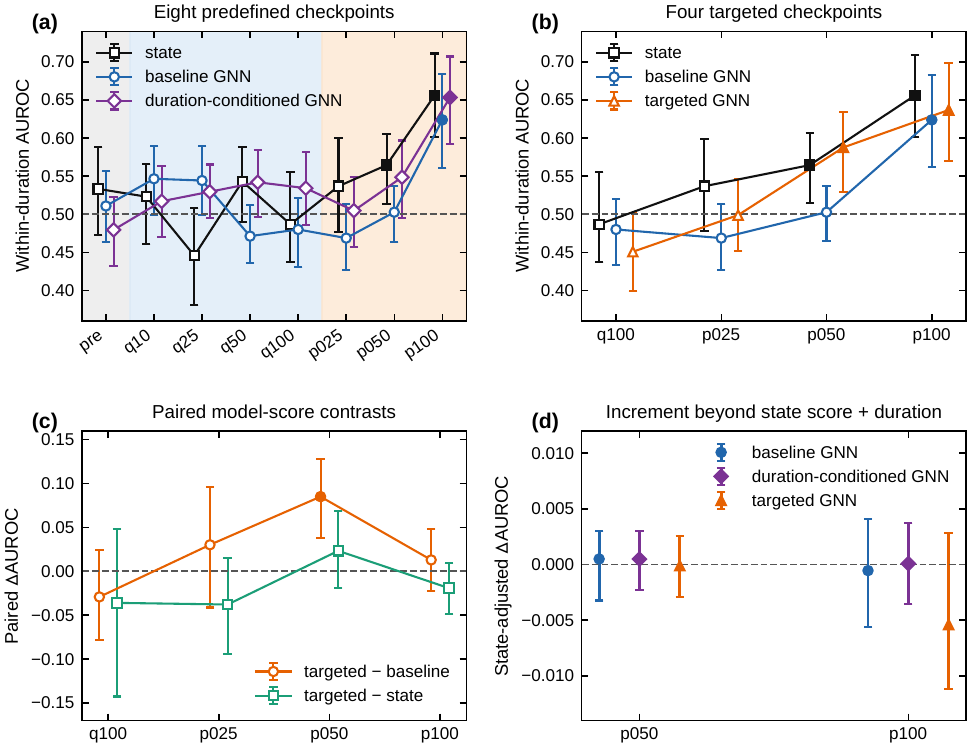}
\caption{\textbf{Exploratory graph-representation tests.} (a) Eight-phase
comparison of the state score, baseline GNN, and duration-conditioned GNN. The
two predefined graph models first pass at p100. (b) Comparison at q100, p025,
p050, and p100 with the targeted GNN sensitivity model; p050 is the first
sampled phase at which both lower limits exceed 0.5.
The gray, blue, and orange backgrounds in panel (a) denote pre-quench,
quench-relative, and post-quench checkpoints, respectively.
(c) Paired targeted-minus-baseline and targeted-minus-state AUROC contrasts.
At p050, these contrasts are +0.085 [0.038, 0.127] and +0.023
[$-0.019$, 0.069], respectively. (d) Cross-fitted state-adjusted AUROC
increments for baseline, duration-conditioned, and targeted GNNs; every
interval includes zero at p050 and p100. The targeted GNN comparison changes
the added channel, history construction, auxiliary losses, and checkpoint
selection together, so it is not a controlled representation ablation. Error
bars are parent-cluster bootstrap intervals. In panels (a--c), filled markers
have hierarchical seed--parent lower limits above 0.5}
\label{fig:s14}
\end{figure}

\begin{figure}[p]
\centering
\includegraphics[width=\textwidth]{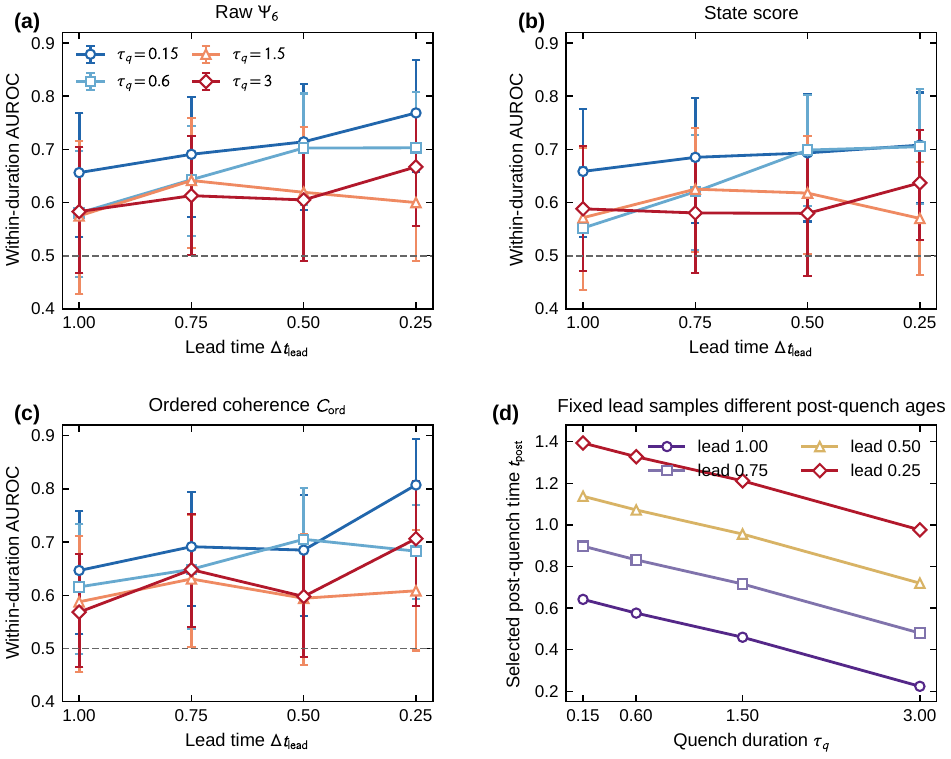}
\caption{\textbf{Quench-duration-specific fixed-lead comparison.} (a--c)
Duration-specific AUROCs for raw $\Psi_6$, the state score,
and $\Cord$ at four leads before the primary terminal window. Error bars are
pointwise parent-cluster 95\% intervals; the corresponding simultaneous
families contain 16 cells per observable and 48 cells jointly. At lead 1.0,
all three pointwise lower limits exceed 0.5 only for $\tau_q=0.15$, and no
duration has this property under the 16-cell correction. Fixed lead maps
to different post-quench ages across protocols. At lead 0.50, all three
16-cell lower limits exceed 0.5 for $\tau_q=0.15$ and 0.60, but none do so for
the two longer durations; Table~\ref{tab:s12couplingduration} adds $\sT$ and
the joint 64-cell family. (d) Selected post-quench time
as a function of quench duration for each requested lead. This deterministic
time-grid map makes the fixed-lead tradeoff explicit}
\label{fig:s15}
\end{figure}

\begin{figure}[p]
\centering
\includegraphics[width=\textwidth]{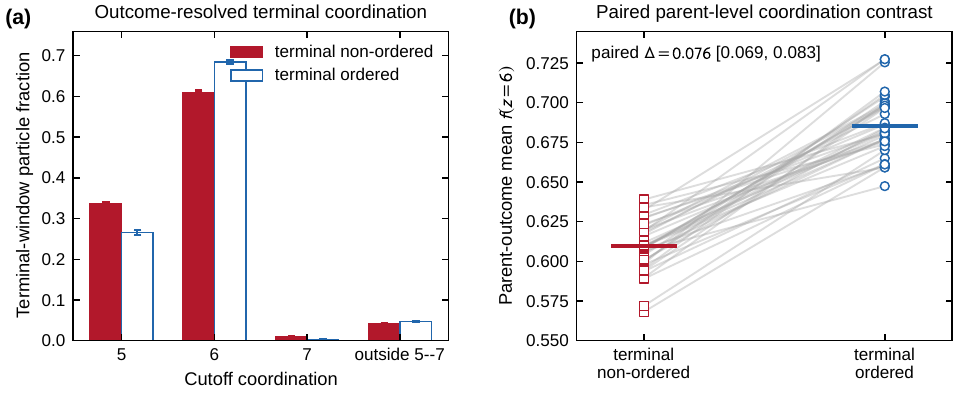}
\caption{\textbf{Terminal coordination and parent-level contrasts.} (a) Full
outcome-resolved terminal cutoff-coordination fractions and parent-cluster
intervals. (b) Parent-by-outcome means of the terminal fraction of
six-coordinated particles. Gray
lines connect the 39 parents contributing both labels; horizontal bars show
equal-parent outcome means. The annotated ordered-minus-non-ordered paired
contrast and interval come from parent-cluster resampling}
\label{fig:s16}
\end{figure}

\begin{figure}[p]
\centering
\includegraphics[width=\textwidth]{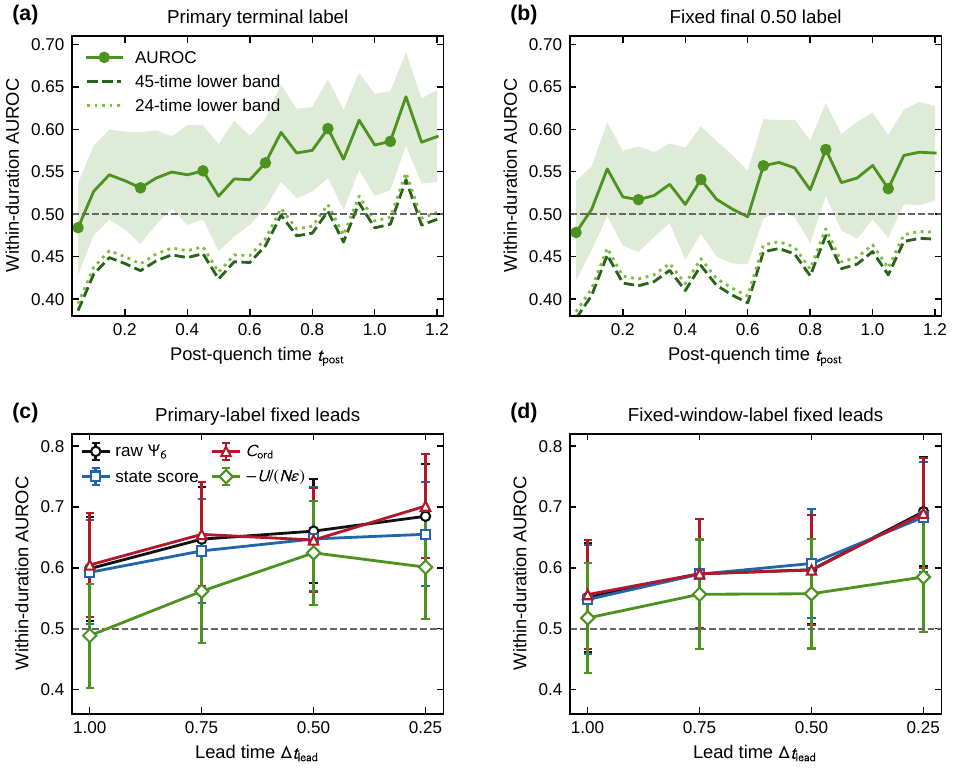}
\caption{\textbf{Non-orientational potential-energy comparison.} (a,b) Dense
within-duration AUROC for $s_E=-U/(N\varepsilon)$ under the primary fractional
and fixed final-0.50 terminal labels. Shading gives pointwise parent-cluster
intervals; dashed and dotted curves are simultaneous lower bands over the full
45-time and 24-time post-quench families. The full 45-time lower band does not
remain above 0.5 for either label through $t_{\rm post}=1.20$; the primary-label
24-time post-quench lower band does so only at the final 1.20 grid point. (c,d)
Fixed-lead AUROCs for raw $\Psi_6$, the state
score, $\Cord$, and $s_E$ under the two labels. Error bars are joint 95\%
parent-cluster intervals over all four observables and four leads. At the
primary-label lead of 1.0, all three orientational observables remain above
chance after correction whereas $s_E$ does not. Under the fixed label, none
has a lower limit above 0.5 at lead 1.0; the three orientational observables,
but not $s_E$, have lower limits above 0.5 at lead 0.50. Under the primary label,
$s_E$ itself has lower limits above 0.5
at leads 0.50 and 0.25, with AUROCs 0.625 and 0.601. The energy comparison is
therefore lead dependent: it does not reproduce the longest-lead or complete
dense-time result, but it is not null near the terminal window. All
observations strictly precede the applicable label window}
\label{fig:s17}
\end{figure}

\begin{figure}[p]
\centering
\includegraphics[width=\textwidth]{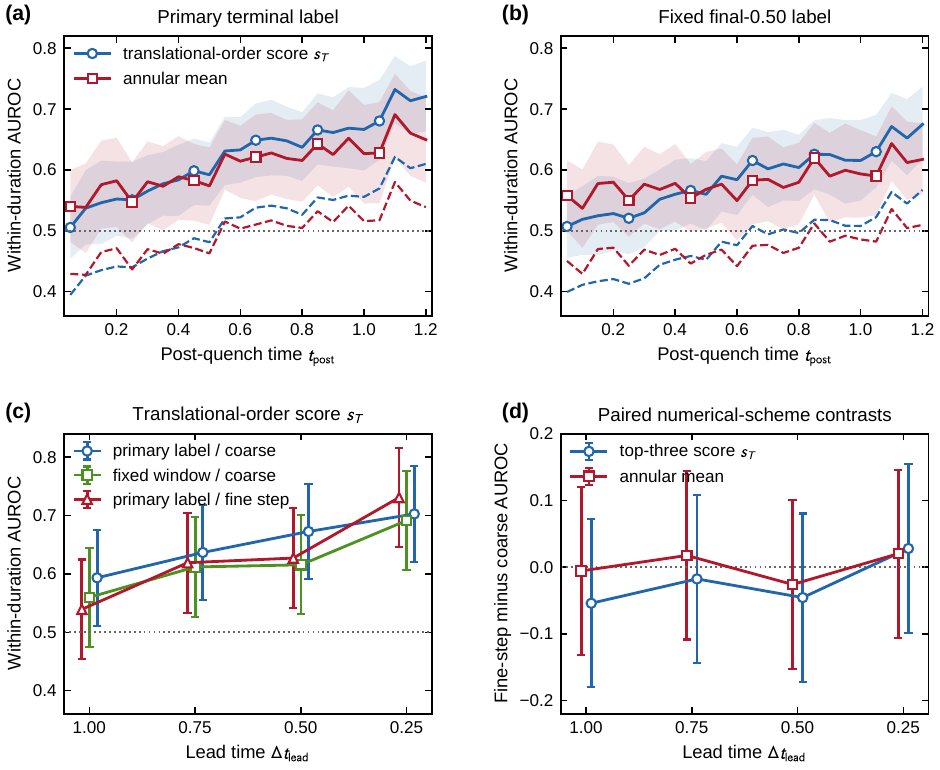}
\caption{\textbf{Translational-order comparison defined without terminal-label information.} (a,b) Dense
within-duration AUROC for the top-three first-reciprocal-shell score $\sT$ and the
annular mean under the primary and fixed-window labels. Shading gives pointwise
95\% parent-cluster intervals; dashed curves are joint lower bands over both
translational scores and all 45 sampled times. (c) Fixed-lead $\sT$ in the
primary, fixed-window-label, and fine-step analyses. Error bars are joint
$4\times4$ intervals over raw $\Psi_6$, the state score, $\Cord$, and $\sT$.
All three lower limits exceed 0.5 at lead 0.50; only the primary-analysis lower
limit exceeds 0.5 at lead 1.0.
(d) Paired fine-step-minus-primary contrasts for both translational scores.
Intervals use the joint $5\times4$ numerical family and all cross zero. The
annular mean is a non-maximizing sensitivity to the top-three summary. The
metric geometry and aggregation do not use the terminal labels. All observations precede the
applicable terminal window}
\label{fig:s18}
\end{figure}

\clearpage
\begin{figure}[H]
\centering
\includegraphics[width=0.95\textwidth]{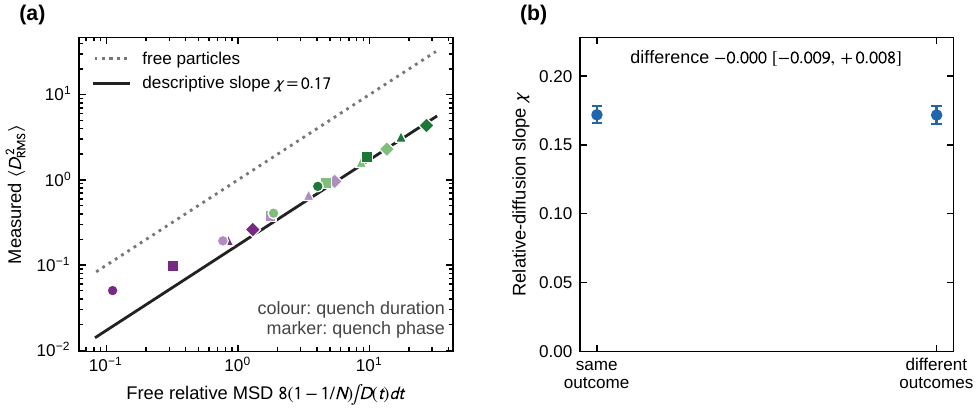}
\caption{\textbf{Replica separation relative to time-integrated diffusivity.}
(a) Translation-corrected mean-squared replica separation, averaged within
each duration-by-checkpoint combination, versus the free-particle reference $X$ in
Eq.~\eqref{eq:s-free-separation}. The line is a descriptive through-origin fit
to all replica pairs; the free-particle relation has unit slope. (b) Fitted
slope $\chi$ for all pairs and for pairs with the same or different terminal
outcomes. Bars show 95\% complete-parent bootstrap intervals. The
different-minus-same interval includes zero, so rapid separation follows the
time-integrated diffusivity similarly for the two pair types}
\label{fig:sexposure}
\end{figure}